\documentclass[apjl]{emulateapj}

\usepackage{natbib}
\usepackage{graphicx}
\usepackage{amsmath,amssymb}
\usepackage{mathptmx}
\usepackage[english]{babel}
\usepackage{booktabs}
\usepackage{multirow}

\def\tvir{\ifmmode {T_{\rm vir}} \else $T_{\rm vir}$\fi}
\def\mvir{\ifmmode {M_{\rm vir}} \else $M_{\rm vir}$\fi}
\def\rvir{\ifmmode {r_{\rm vir}} \else $r_{\rm vir}$\fi}
\def\rcgm{\ifmmode {r_{\rm CGM}} \else $r_{\rm CGM}$\fi}
\def\mdm{\ifmmode {M_{\rm DM}} \else $M_{\rm DM}$\fi}
\def\mdisk{\ifmmode {M_{\rm disk}} \else $M_{\rm disk}$\fi}
\def\mb{\ifmmode {M_{\rm b}} \else $M_{\rm b}$\fi}
\def\mcor{\ifmmode {M_{\rm corona}} \else $M_{\rm corona}$\fi}
\def\mbm{\ifmmode {M_{\rm b,miss}} \else $M_{\rm b,miss}$\fi}
\def\mbo{\ifmmode {M_{\rm b,obs}} \else $M_{\rm b,obs}$\fi}
\def\fb{\ifmmode {f_{\rm b}} \else $f_{\rm b}$\fi}

\def\tgas{\ifmmode {T_{\rm gas}} \else $T_{\rm gas}$\fi}
\def\rgas{\ifmmode {r_{\rm gas}} \else $r_{\rm gas}$\fi}
\def\rogas{\ifmmode {\rho_{\rm gas}} \else $\rho_{\rm gas}$\fi}
\def\mgas{\ifmmode {M_{\rm WM}} \else $M_{\rm WM}$\fi}
\def\mhi{\ifmmode {M_{HI}} \else $M_{HI}$\fi}
\def\nhi{\ifmmode {N_{HI}} \else $N_{HI}$\fi}
\def\rhi{\ifmmode {R_{\rm HI}} \else $R_{\rm HI}$\fi}
\def\rhalf{\ifmmode {r_{1/2}} \else $r_{1/2}$\fi}
\def\npeak{\ifmmode {N_{\rm peak}} \else $N_{\rm peak}$\fi}
\def\tdyn{\ifmmode {t_{\rm dyn}} \else $t_{\rm dyn}$\fi}
\def\tcool{\ifmmode {t_{\rm cool}} \else $t_{\rm cool}$\fi}
\def\fcool{\ifmmode {f_{\rm cool}} \else $f_{\rm cool}$\fi}
\def\trec{\ifmmode {t_{\rm rec}} \else $t_{\rm rec}$\fi}
\def\Tmin{\ifmmode {T_{\rm min}} \else $T_{\rm min}$\fi}

\def\kal{\ifmmode {K_{\alpha}} \else $K_{\alpha}$\fi}
\def\kbe{\ifmmode {K_{\beta}} \else $K_{\beta}$\fi}

\def\nh{\ifmmode {n_{\rm H}} \else $n_{\rm H}$\fi}
\def\ca{\ifmmode {\chi_{\rm EM}} \else $\chi_{\rm EM}$\fi}
\def\cb{\ifmmode {\chi_{\rm N}} \else $\chi_{\rm N}$\fi}
\def\lia{\ifmmode {22~{\rm \AA}} \else $22~{\rm \AA}$\fi}
\def\lib{\ifmmode {19~{\rm \AA}} \else $19~{\rm \AA}$\fi}

\def\cg{\ifmmode {c_g} \else $c_g$\fi}
\def\c6{\ifmmode {c_{g,6}} \else $c_{g,6}$\fi}
\def\phim{\ifmmode {P_{\rm HIM}} \else $P_{\rm HIM}$\fi}

\def\kb{\ifmmode {k_{\rm B}} \else $k_{\rm B}$\fi}
\def\mp{\ifmmode {m_{p}} \else $m_{p}$\fi}

\def\chisq{\ifmmode {\chi_{mod}^2} \else $\chi_{mod}^2$\fi}

\def\msun{\ifmmode {\rm M_{\odot}} \else $\rm M_{\odot}$\fi}
\def\lsun{\ifmmode {\rm L_{\odot}} \else $\rm L_{\odot}$\fi}

\def\kms{\ifmmode {\rm km\:s^{-1}} \else $\rm km\:s^{-1}$\fi}
\def\cmc{\ifmmode {\rm cm^{-2}} \else $\rm cm^{-2}$\fi}
\def\cmv{\ifmmode {\rm cm^{-3} \:} \else $\rm cm^{-3}$\fi}
\def\kel{\ifmmode {\rm \:\: K} \else $\rm \:\: K$\fi}
\def\pc{\ifmmode {\rm \:\: pc} \else $\rm \:\: pc$\fi}
\def\kpc{\ifmmode {\rm \:\: kpc} \else $\rm \:\: kpc$\fi}
\def\amu{\ifmmode {\rm \:\: amu} \else $\rm \:\: amu$\fi}
\def\fluxun{\ifmmode {\rm \:\: erg~s^{-1}~cm^{-2}~deg^{-2}} \else $\rm \:\: erg~s^{-1}~cm^{-2}~deg^{-2}$\fi}
\def\liun{\ifmmode {\rm \:\: photons~s^{-1}~cm^{-2}~sr^{-1}} \else $\rm \:\: photons~s^{-1}~cm^{-2}~sr^{-1}$\fi}

\def\msuny{\ifmmode {\rm M_{\odot}\:year^{-1}} \else $\rm M_{\odot}\:year^{-1}$\fi}
\def\ergs{\ifmmode {\rm erg\:s^{-1}} \else $\rm erg\:s^{-1}$\fi}

\def\ao{\ifmmode {a_{\rm O}} \else $a_{\rm O}$\fi}

\def\gv{\ifmmode {g_{\mbox{\tiny {\it V}}}} \else $g_{\mbox{\tiny {\it V}}}$\fi}
\def\gM{\ifmmode {g_{\mbox{\tiny {\it M}}}} \else $g_{\mbox{\tiny {\it M}}}$\fi}

\def\ppv{\ifmmode {p_{\mbox{\tiny {\it V}}}} \else $p_{\mbox{\tiny {\it V}}}$\fi}
\def\ppm{\ifmmode {p_{\mbox{\tiny {\it M}}}} \else $p_{\mbox{\tiny {\it M}}}$\fi}
\def\pfv{\ifmmode {f_{\mbox{\tiny {\it V}}}} \else $f_{\mbox{\tiny {\it V}}}$\fi}
\def\pfm{\ifmmode {f_{\mbox{\tiny {\it M}}}} \else $f_{\mbox{\tiny {\it M}}}$\fi}

\def\pvw{\ifmmode {p_{\mbox{\tiny {\it V}},w}} \else $p_{\mbox{\tiny {\it V}},w}$\fi}
\def\pmw{\ifmmode {p_{\mbox{\tiny {\it M}},w}} \else $p_{\mbox{\tiny {\it M}},w}$\fi}
\def\fvw{\ifmmode {f_{\mbox{\tiny {\it V}},w}} \else $f_{\mbox{\tiny {\it V}},w}$\fi}
\def\fmw{\ifmmode {f_{\mbox{\tiny {\it M}},w}} \else $f_{\mbox{\tiny {\it M}},w}$\fi}

\def\FSM17{Faerman~et~al.~(2017)}
\def\GFS17{Gottlieb~et~al.~(2017)}

\shorttitle{Massive Coronae}
\shortauthors{Faerman et al.}
\slugcomment{Accepted to ApJ}

\begin{document}

\title{Massive Warm/Hot Galaxy Coronae as Probed by UV/X-ray Oxygen Absorption and Emission: \\ I - Basic Model}

\author{
Yakov Faerman, \altaffilmark{1}
Amiel Sternberg\altaffilmark{1}
Christopher F. McKee \altaffilmark{2}
}

\altaffiltext{1}
{Raymond and Beverly Sackler School of Physics and Astronomy,
Tel Aviv University, Ramat Aviv 69978, Israel; yakovfae@post.tau.ac.il}
\altaffiltext{2}
{Department of Physics and Department of Astronomy,
University of California at Berkeley, Berkeley CA 94720}

\email{yakovfae@post.tau.ac.il}

\begin{abstract}
We construct an analytic phenomenological model for extended warm/hot gaseous coronae of $L_*$ galaxies. We consider UV OVI COS-Halos absorption line data in combination with Milky Way X-ray OVII and OVIII absorption and emission. We fit these data with a single model representing the COS-Halos galaxies and a Galactic corona. Our model is multi-phased, with hot and warm gas components, each with a (turbulent) log-normal distribution of temperatures and densities. The hot gas, traced by the X-ray absorption and emission, is in hydrostatic equilibrium in a Milky Way gravitational potential. The median temperature of the hot gas is $1.5 \times 10^6$~K and the mean hydrogen density is $\sim 5 \times 10^{-5}$~\cmv. The warm component as traced by the OVI, is gas that has cooled out of the high density tail of the hot component. The total warm/hot gas mass is high and is $1.2 \times 10^{11}$~\msun. The gas metallicity we require to reproduce the oxygen ion column densities is $0.5$ solar. The warm OVI component has a short cooling time ($\sim 2 \times 10^8$~years), as hinted by observations. The hot component, however, is $\sim 80\%$ of the total gas mass and is relatively long-lived, with $\tcool \sim 7 \times 10^{9}$~years. Our model supports suggestions that hot galactic coronae can contain significant amounts of gas. These reservoirs may enable galaxies to continue forming stars steadily for long periods of time and account for ``missing baryons" in galaxies in the local universe.
\end{abstract}

\keywords{galaxies: formation --- galaxies: halos --- intergalactic medium --- quasars: absorption lines --- X-ray: galaxies --- UV:galaxies}

%
%

\section{Introduction}
\label{sec_intro}

Ultraviolet and X-ray studies have provided convincing observational evidence for the existence of hot, extended gaseous coronae around star-forming galaxies. This includes XMM-Newton and Chandra detections of local (redshift $z \sim 0$) OVII and OVIII emission data \citep{Henley10,Henley10lines} and absorption lines towards bright QSOs \citep{Nicastro02,Rasmussen03}, suggesting high column densities of highly ionized oxygen in the vicinity of the Milky Way (MW) \citep{Fang06,Bregman07}. More recent are Cosmic Origins Spectrograph (COS) ultraviolet observations of OVI absorbers at large distances, up to at least $150$~kpc, in the circumgalactic medium (CGM) around many galaxies \citep[hereafter T11]{Tumlinson11}. Lower ionization species such as CII, SiIII and NV, are also detected in the CGM of many of the COS-Halos survey galaxies (\citealp{Werk13, Stocke13, Werk16}).  Our focus in this paper is the highly ionized oxygen.

Additional evidence for hot coronae is the X-ray line and continuum emission observed in local universe galaxies, extending beyond the stellar disks \citep{Anderson11,Bogdan13,AGW15}. Furthermore, a hot medium around our galaxy is necessary to explain gas stripping in MW satellites \citep{Blitz00,Grcevich09,Gatto13}. Extended hot coronae are also invoked and found by analytical calculations \citep{Spitzer56, WR78,White91,Fukugita06} and hydrodynamics simulations \citep{Cen99,Ntormousi10,Crain10,Joung12b,Cen13,Liang16,RocaF16}.

We know from cosmological measurements that ``baryons" (i.e.~normal atoms) consitute $ֿ15.7\%$ of the matter in the universe by mass \citep{Planck16}. However, for galaxies in the local universe only a fraction of the expected baryonic mass is detected, with $30-70\%$  missing \citep{Silk03,Fukugita04,Prochaska09,Shull12}. Hot coronae may account for some of these missing baryons in galaxies~\citep{MB04, Fukugita06}.

Additional motivation for studying hot coronae is a better understanding of galaxy formation and evolution processes. How do galaxies acquire their gas and is star formation regulated by external reservoirs? How do hot coronae interact with other parts of galaxy structures, especially at disk boundaries?

One plausible scenario for the formation of hot coronae is gas accretion from the intergalactic medium (IGM). As the gas falls into the gravitational potential of the dark matter halos a shock is formed near the virial radius \citep{Birnboim03,Keres05,Cen06a}. The shock heats the gas to the virial temperature. Simulations show that the post-shock gas is decelerated and a (quasi-)static corona is formed (e.g.~\citealt{Nelson16}). Alternatively, coronae may be produced by supernovae- (SNe) and AGN-driven massive winds \citep{Veil05, Sturm11, Genzel14}, extending to the virial radius of the galaxy and beyond \citep{Sokol16}. The winds deposit energy and momentum into the CGM and enrich it with metals \citep{Genel14, Suresh15}. The ejected gas may re-accrete onto the galaxy in a large scale version of the ``galactic fountain" \citep{Shapiro76,Bregman80,Breitschwerdt00,Marinacci2011}.

For our Galaxy, several attempts have been made to model a hot corona and its properties using OVII absorption lines, with the implied total hot gas mass as an interesting but debated result. For example, \citet{Anderson10} assume the hot gas in the MW follows the NFW dark matter distribution and conclude that the total gas mass is low, only a few percent of the cosmological fraction. More recently, \citet{Miller13} fit an isothermal $\beta$-model (used originally to fit surface brightness profiles of galaxy clusters, see \citealp{Forman85} and \citealp{OPC03}) to a similar data set with similar conclusions. \citet{Fang13} suggest that while a cuspy corona can only contain a small gas mass, a galaxy with a low density extended corona could be ``baryonically closed", with the baryonic to total mass ratio equal to the cosmological value (see also \citealp{MB04} and \citealp{Sommer06}). A simplified analysis by \citet{Gupta12} also shows that the gas mass in the corona may be significant and contain all the missing Galactic baryons.
\cite{Tepper15} construct a corona model in which the gas is in hydrostatic equilibrium in the potential of the galaxy and simulate its interaction with the Magelanic Stream, constraining the hot gas density at large radii.

In this paper we combine the observations of extended OVI absorbers in the COS-Halos galaxies together with the OVII and OVIII absorption and emission associated with our Galaxy in a single, Milky-Way-based, unified model for the typical coronae of $L^*$ galaxies. Combining the OVI with OVII and OVIII observations is challenging since their simultaneous presence implies the existence of gas at a wide range of temperatures, from $2 \times 10^5$~K to $3 \times 10^6$~K (see also \citealp{Furl05}). The main goal of this paper is to present the basic features of our model and a fiducial computation that accounts for the observational constraints. We will present a parameter study in paper II of this series.

Our model is phenomenological and is based on simple physical assumptions regarding the gas distributions. The important advances compared to previous works are the combination of UV and X-ray data in a single model and including the effects of turbulence on the gas properties and line widths. In \S \ref{sec_obs} we present observational evidence for the hot corona and discuss its main properties in \S \ref{sec_motivation}. We present the basic ingredients of our model and its output properties in \S \ref{sec_model} and \S \ref{sec_results}. We discuss the implications of our results in \S \ref{sec_discussion}, and summarize in \S \ref{sec_summary}.

%

\section{Observational Constraints}
\label{sec_obs}

In this section we present the data we aim to reproduce. First, we examine the oxygen ion column densities, from OVI to OVIII, as inferred from observed line absorptions towards QSOs. In our analysis we combine several different data sets. In the X-ray, we consider local ($z \sim 0$) absorption, OVII and OVIII lines, originating in the vicinity of the Milky Way. In the UV, we examine the OVI absorption line and consider data collected in the COS-Halos survey, probing the CGM of $\sim L^*$ galaxies at $z \sim 0.3$. In addition to the OVI, OVII and OVIII absorption data, we also consider observations of line and broad-band soft X-ray emission from the MW halo, as well as pulsar dispersion measure observations.

The reason to examine OVI data from the COS-Halos galaxies is that MW OVI is heterogenous with significant contributions from nearby low-velocity gas in addition to extended circumgalactic material (see \citealp{Savage03,Sembach03}). \cite{Zheng15} argue that as much as half of the circumgalactic medium (CGM) of the Milky Way as probed by OVI is hidden by nearby, low velocity gas. The COS-Halos data probe the extended CGM of MW-sized galaxies, at distances of up to 150 kpc from the galaxies. As we discuss below, the MW is also similar to the COS-Halos galaxies in terms of its star formation rate and, after the correction suggested by \cite{Zheng15}, its OVI column. This motivates us to combine data from the MW and the COS-Halos galaxies in our model. Furthermore, we use the OVI COS-Halos data to construct a unified spatial profile, probing the OVI distribution around a single effective galaxy, as an MW model. 

\subsection{OVI absorbers}
\label{subsec_obs_ovi}

\begin{figure}
\includegraphics[width=0.45\textwidth]{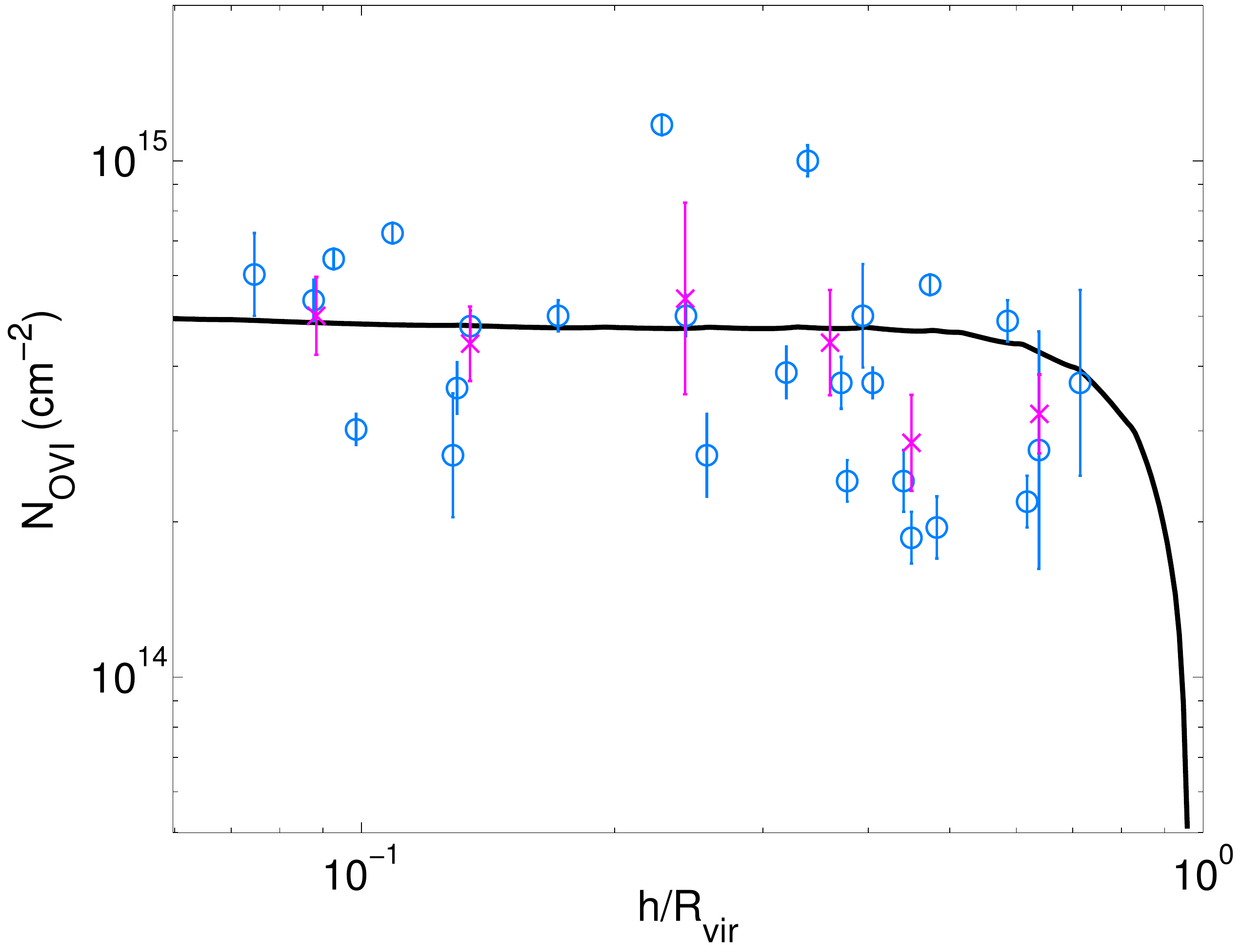}
\caption{Observed OVI column densities and model fit. The blue crosses show the original measurements, from \citet{Tumlinson11} and \citet{Werk13}, as a function of the impact parameter, $h$, normalized by the virial radius for each galaxy. Magenta shows the data after binning (see \S \ref{sec_obs} for details). The black solid line shows the result of our fiducial model (see \S \ref{sec_results}).}
   \label{fig:ovi}
 \end{figure}

In the UV, we consider the HST/COS data set from T11, measuring OVI absorption ($\lambda\lambda~1031.9,1037.6 ~\AA$). These are spectra of QSO sightlines probing the circumgalactic medium of $\sim~30$ $L^*$ galaxies at redshifts of $0.1$ to $0.4$, at impact parameters of up to 150 kpc from the centers of the galaxies. These detections clearly indicate the existence of extended coronal gas structures around galaxies. Since the galaxies are similar in their properties, we attempt to treat these lines of sight (typically, one per galaxy) as probing a single effective corona, but at different impact parameters. For this analysis we account for variations in galaxy size and dark matter halo properties by normalizing the measured impact parameters to the virial radii (as estimated by \citealp{Werk13}). The OVI columns for the individual sightlines are the blue points in Figure \ref{fig:ovi}. We reduce the scatter by binning the data as indicated by the magenta points and error bars. Our model will then fit for the radial distribution of the binned data. The observed OVI column densities are distributed fairly evenly to large radii, suggesting flat gas density profiles inside the dark matter halos. The solid curve in the figure is our model fit, which we describe in \S \ref{sec_model} and \S \ref{sec_results}.

The OVI absorption lines are spectrally resolved, and their properties as measured by \citet{Werk13} offer some hints of the gas kinematics. First, the line widths, with $\sigma \sim 40$~\kms, are larger than expected for gas at $T \sim 3 \times 10^5$~K, at which the OVI is expected to be abundant assuming collisional ionization equilibrium (CIE). 
Furthermore, the central velocities of the OVI lines typically deviate from the galactic velocities by $50$ to $100$~\kms. These two results suggest that non-thermal motions are dynamically significant. Possible explanations for the observed velocities include rotational motions and local inflows and outflows.
The standard deviation of the mean line velocities relative to the host galaxies is $\sim 70$~\kms. We use this as a measure of the turbulent velocity scale and consider $\sigma_{\rm turb}$ in the range of $60-80$~\kms. This estimate is consistent with the OVII and OVIII line width analysis we present next (\S \ref{subsec_obs_ovii}). If other sources of motion are significant and contribute both to line shift and broadening, this velocity is an upper limit on the turbulent velocity scale.

The star-forming galaxies of the COS-Halos sample are similar in their luminosities and star formation rates to the Milky Way. T11 note that the typical OVI column in their sample, with $N_{\rm OVI} \sim 3-4 \times 10^{14}~\cmc$, is higher than the mean value measured for the MW, $\sim 10^{14}~\cmc$, in FUSE observations of high velocity gas (see \citealp{Sembach03, Savage03}). However, \cite{Zheng15} show that a significant fraction of the MW CGM may have low velocities, making it difficult to separate from warm/hot gas in the disk using only velocity cuts. In their simulations, half of the total OVI absorption originating in the CGM remains `hidden' in MW observations. \cite{Zheng15} correct for this, increasing the MW OVI column density to $\sim 2 \times 10^{14}~\cmc$, bringing it closer to to the OVI column densities measured in the COS-Halos galaxies. We note that, with a specific star formation rate (sSFR) of $\sim 3-4 \times 10^{-11}~\msuny$ (\citealp{Chomiuk11,Lic15}), the MW is consistent with the low end of the $N_{\rm OVI}$-sSFR correlation found by T11 for their star-forming galaxies (see their Figure 3).

As we explain in \S\ref{sec_model}, in our model we assume that the coronal gas is in CIE. Another possible production mechanism for the OVI is photoionization by metagalactic radiation, but as we discuss in \S \ref{subsec_disc_photo} this is insignificant in our model.

\subsection{OVII \& OVIII absorbers}
\label{subsec_obs_ovii}

In the X-ray, we focus on the local OVII and OVIII absorption lines, produced in gas associated with our Galaxy.
We consider the observational data and estimate typical values for the column densities, which we then fit with our model.
Throughout section \S \ref{sec_obs}, we use the median as the typical value of a given sample and estimate the 1-$\sigma$~errors as the standard deviation of the sample reduced by a factor of $\sqrt{N-1}$, where $N$ is the sample size.

Since X-ray absorption lines are not fully resolved even with grating spectrometers, the basic analysis is based on the line equivalent width (EW), and the inferred column density depends on the assumed line velocity width. \citet[hereafter MB13]{Miller13} measure the EW of the OVII~\kal~line ($\lambda=21.60~\AA$) for a set of $29$ sightlines. They assume a Doppler parameter of $150$~\kms, motivated by the thermal velocity of hydrogen gas at $2 \times 10^6$~K, resulting in slightly saturated lines. 

\citet[hereafter F15]{Fang15} analyze a larger data set, of $43$ sightlines, detect OVII absorption in 33 objects, and fit the observed spectra to derive both the line widths and column densities. For the column densities, the sightlines with detected absorption have a median value of $3.7~\times10^{16}$~\cmc, with with a 1 $\sigma$ error range of $2.6-5.4 \times10^{16}$~\cmc. The constraints on the Doppler parameter are not very strong for most of these objects. However, assuming the absorption lines have a common velocity width, we estimate a median value (and error) of $b=98~(79-117)$~\kms. For $10$ objects in their sample, F15 do not detect absorption and derive upper limits for the EW. We translate these into upper limits for the column densities, assuming the typical Doppler parameter of the detected lines. Adding the upper limits to the group of objects in which absorption is detected decreases the median value to $N_{\rm OVII}=1.4~(1.0-2.0) \times 10^{16}$~\cmc.

\citet[hereafter G12]{Gupta12} present a smaller sample of 8 bright QSOs for which they are able to measure the OVII \kal, \kbe~($\lambda=21.60$~and~$18.63~\AA$) and OVIII~\kal~($18.97~\AA$) lines. Given two absorption lines of the same ion and assuming a common origin for the absorption, it is possible to constrain the line width and column density using the curve of growth (see G12 and Figure 3 in \citealt{Williams05}). F15 compare equivalent width measurements performed for the same sightlines by different groups and find that \citet{Williams05} underestimate the EW by $\sim 30\%$ (see \citealp{Rasmussen07} for details). Since G12 present the same OVII \kal~EW value for the Mrk 421 sightline as measured by \citet{Williams05}, we conclude that their measurements suffer from similar systematics. We use the EWs measured by G12, increase them by $30\%$, and repeat their joint analysis of the~\kal~and~\kbe~OVII lines. The resulting median values (and 1~$\sigma$ errors) are $b=95~(78-112)~\kms$ and $N_{\rm OVII}= 2.2~(1.8-2.9) \times 10^{16}$~\cmc.

Assuming the OVII and OVIII originate in the same gas and the line Doppler parameter is the same for both ions, we apply the line width derived from OVII to the OVIII EWs from G12 (also increased by $30\%$) to infer the OVIII column densities and estimate a typical column of $N_{\rm OVIII} = 1.0~(0.76-1.3) \times 10^{16}$~\cmc. In CIE, the ratio of OVII to OVIII is very sensitive to the gas temperature and can be a useful constraint. We calculate the ratio of OVII to OVIII columns for the G12 objects and find a median value (and error) of $ 4.0~(2.8-5.6)$, corresponding to $T \sim 1.65 \times 10^6$~K.

The median OVII column density estimated for the G12 sample is higher than that of the F15 sample. A possible explanation is that since G12 focus on spectra with the highest S/N ratio detections, these are the sightlines that show the strongest absorption and highest column densities. The F15 sample, on the other hand, includes detections with lower SNR and upper limits from non-detections. The OVIII column density inferred from G12 may suffer from a similar bias, overestimating the $N_{\rm OVIII}$ typical for the MW corona. However, the OVII/OVIII ratio probes the gas temperature and is independent of the absolute column density values. To obtain a better estimate of the typical OVIII column, we take the F15-derived OVII column density and divide it by the OVII/OVIII typical ratio calculated from the G12 measurements. The resulting column density is $N_{\rm OVIII} =  0.36~(0.22-0.57)\times 10^{16}$~\cmc.

\bgroup
\def\arraystretch{1.5}

\begin{table*}
\centering
	\caption{Summary of Milky Way X-ray observational data}
	\label{tab:xray_obs}
		\begin{tabular} {| l | c | c |}
			\toprule
\multicolumn{1}{|c|}{Data set} & Quantity & Typical value (1-$\sigma$~error range) \\
			\midrule
\multicolumn{3}{|c|}{Absorption~(see \S \ref{subsec_obs_ovii} for details)}			\\
			\midrule
\multirow{4}{*}{\citet{Gupta12}}	& $b$ 		 									& $95.0~(77.9-112.1)$~\kms  \\ 
													& $N_{\rm OVII}$ \footnote{~Absorption detected in all sightlines.\label{ftn:det1}}
													& $2.2~(1.8-2.9) \times 10^{16}$~\cmc	\\
													& 	$N_{\rm OVIII} ^{\rm \ref{ftn:det1}}$						
													& $1.0~(0.76-1.3) \times 10^{16}$~\cmc	 \\
													& $\rm OVII/OVIII$~ratio				& $4.0~(2.8-5.6)$ \\
			\midrule
\multirow{2}{*}{\citet{Fang15}} 	& $b$ 											& $98.0~(78.6-117.4)$ ~\kms  \\ 
			& $N_{\rm OVII}$ \footnote{~Including upper limits from non-detections.}
												 	& $1.4~(1.0-2.0) \times 10^{16}$~\cmc	\\
			\midrule
			---  & $N_{\rm OVIII}$ 
			\footnote{~Using the \cite{Fang15} OVII and dividing by the \cite{Gupta12} OVII/OVIII ratio (see \S \ref{subsec_obs_ovii}).}
					&  $0.36~(0.22-0.57) \times 10^{16}$~\cmc \\
			\midrule
\multicolumn{3}{|c|}{Emission~(see \S \ref{subsec_obs_emission})}			\\
			\midrule
\citet{Henley10} \footnote{~Emission detected in all observed fields.\label{ftn:det2}}
					& $S_{0.4-2.0} $		& $2.1~(1.9-2.4) \times 10^{-12}~\fluxun$  \\
			\hline
\multirow{3}{*}{\citet{Henley10lines} $^{\rm \ref{ftn:det2}}$} 		& $22~\rm \AA$	 
							\footnote{After foreground correction (see \citealp{Henley10lines}, Figure 18).\label{ftn:corr}} 
					& $2.8~(2.3-3.4)~\liun$ \\
					& $19~\rm \AA$	$^{\rm \ref{ftn:corr}}$	& $0.69~(0.58-0.83)~\liun$ \\
					& $\lia/\lib$~ratio	& $4.3~(3.4-5.5) $ \\
			\bottomrule			
		\end{tabular}
\end{table*}

We summarize the X-ray absorption observations we adopt in this work in Table~\ref{tab:xray_obs}. For the line width and the OVII column density we adopt the values derived from the full sample in F15 (including upper limits), $b = 98~(79-117)$~\kms~and $N_{\rm OVII} = 1.4~(1.0-2.0) \times 10^{16}$~\cmc. We choose this sample since it is the largest published collection of OVII absorption lines and the analysis includes estimation of the line width. We note that the Doppler parameters we derive from G12 and F15 are determined independently and are consistent with each other. Importantly, they are also consistent with an oxygen line width in gas at $2 \times 10^6$~K (a typical temperature for the OVII and OVIII ions) with a turbulent component of $b_{\rm turb} = \sqrt{2} \sigma_{\rm turb} \sim 85$~\kms, similar to the value we estimated from the T11 OVI data. For the OVIII column density, we adopt $N_{\rm OVIII} = 0.36~(0.22-0.57) \times 10^{16}$~\cmc, derived from the F15 OVII column and the G12 OVII/OVIII ratio, $4.0~(2.8-5.6)$.

In our analysis, we will be assuming that all of the OVII and OVIII absorptions occur in the hot corona, and in this paper we do not attempt to correct for possible contributions from the Galactic disk. Thus, for these high oxygen ions, the observed column densities are in fact upper limits for our model, which includes only coronal gas.

\subsection{X-ray emission}
\label{subsec_obs_emission}

OVII and OVIII line and broad band X-ray emission also provide important constraints, as direct probes of the gas emission measures. We consider observations of 0.4-2.0 keV soft X-ray emission presented by \cite{Henley10}, and \lia~and \lib~line emissions presented by \cite{Henley10lines}, and attributed to hot halo gas after corrections for foregrounds. The emission observations we adopt as constraints for our model are summarized in Table~\ref{tab:xray_obs}.

\subsubsection{Line Emission}
\label{subsub_lines}

The helium-like OVII ion and the hydrogen-like OVIII ion produce line emission features near $\sim \lia$~(0.56 keV) and \lib~(0.65 keV). The \lia~feature is due entirely to OVII and is a blend of the singlet $21.6~\AA$ resonance transition and the triplet-to-singlet intercombination and forbidden transitions at $22.1$ and $21.8~\AA$. This complex of lines is generally not resolved by the X-ray CCD cameras \citep{Snowden04}.

For a gas containing a mixture of OVII and OVIII, the \lib~feature is a blend of OVIII Ly$\alpha$ and an OVII intercombination line at $18.63~\AA$. For collisionally excited gas in CIE, the relative contributions of OVII and OVIII to the \lib~feature depend sensitively on the temperature. Table~\ref{tab:xray_atomic} presents a summary of the transitions contributing to the \lia~and \lib~features (see also \citealp{Aggarwal08}).

\citet[hereafter HS10]{Henley10lines} present measurements of the \lia~and \lib~emission intensities. They discuss and deal with contamination from solar-wind charge exchange (SWCX), bright X-ray sources in the observed fields and the extragalactic X-ray background. Furthermore, they correct for the foreground X-ray emission and absorption. The HS10 final results for the corrected oxygen line intensities are presented in their Figure 18, and these are the data we consider here. For the \lia~line intensity, the measured values are in the range 0.$08-5.51$~\liun~(L.U.), and for the \lib~line - $0.10-2.53$~L.U. The typical intensities (and 1-$\sigma$ errors) are $2.8~(2.3-3.4)$ and $0.69~(0.58-0.83)$~\liun, for the \lia~and \lib, respectively. We also compute the \lia/\lib~intensity ratio for each sightline as an observational constraint of the hot gas temperature. The median of the ratios is $4.3~(3.4-5.5)$, corresponding to $T \sim 1.60 \times 10^6$~K.

\bgroup
\def\arraystretch{2}

\begin{table}
\centering
	\caption{Oxygen Transitions in the X-ray}
	\label{tab:xray_atomic}
		\begin{tabular} {| c | c | c | c | c |}
		\toprule
Feature 						&  Ion   	&  Transition	  & $\lambda ~(\rm \AA)$	&  E (eV) 	\\ 
		\midrule
		\midrule
\multirow{3}{*}{\lia}		&	\multirow{3}{*}{O VII}  	&  $2^1P~\rightarrow~1^1S$			& $21.60$		&  573.9 		\\ 
																		&	&  $2^3P~\rightarrow~1^1S$ 			& $22.10$		&  561.0 		\\ 
																		&	&  $2^3S~\rightarrow~1^1S$ 			& $21.80$		&  568.6 		\\ 
		\midrule
\multirow{2}{*}{\lib}		&	O VII  	&  $3^1P~\rightarrow~1^1S$ 				& $18.63$				&  665.6 		\\ 
									&	O VIII   	&  $2^2P~\rightarrow~1^2S$ 				& $18.97$				&  653.6 		\\ 
			\bottomrule			
		\end{tabular}
\end{table}

\subsubsection{Emission in 0.4-2.0 keV}
\label{subsub_band}

\citet[hereafter H10]{Henley10} present X-ray emission observations of 26 fields at high Galactic latitudes and away from the Galactic Center (see their Figure 1). In their analysis, H10 model and subtract the contributions of the local foreground and the extragalactic background to the observed X-ray spectrum, aiming to identify the emission from the hot halo gas. We consider the flux measurements in the 0.4-2.0 keV band for these fields, summarized in Table~2 of H10. The measured fluxes are in the range of $0.51-5.48$~\fluxun, with a median value of $2.1~(1.9-2.4) \times 10^{-12}$~\fluxun.

Typically $\sim 50\%$ of the $0.4-2.0$~keV flux is due to the OVII and OVIII emission lines discussed above.

\subsubsection{External Galaxies}
\label{subsec_obs_ext}
 
An additional direct observational constraint comes from detection of extended X-ray emission around massive spirals in the local universe.
\citet{Rasmussen09} summarizes existing observations in the 0.3-2.0 keV band out to several tens of kpc around several galaxies \citep{Strickland04a, Tullman06a}. The X-ray luminosities are between $10^{39}$ and $10^{40}$~\ergs, with a typical value of $5 \times 10^{39}$~\ergs. \citet{Anderson11} and \citet{Bogdan13} detect diffuse X-ray emission around NGC~1961 and NGC~6753. 
For NGC~1961, \citet{Anderson11} find a luminosity of $\sim 4 \times 10^{40}$~\ergs~between 0.6 and 2.0 keV within 50~kpc of the galaxy. This was revised by \citet{Anderson15} to $\sim 9 \times 10^{40}$~\ergs. \citet{Bogdan13} estimate a bolometric luminosity of $6 \times 10^{40}$~\ergs~inside an annulus of $23-70$~kpc for NGC~1961 and NGC~6753. However, since these galaxies are much more massive, with stellar masses an order of magnitude higher than the MW, they serve as upper limits for our current work.

\subsection{Dispersion Measure}
\label{subsec_obs_other}

Finally, the observed dispersion measure (DM) can also be employed to constrain models of the corona structure. \citet{Anderson10} use pulsar observations by \citet{Manchester06} to estimate a lower limit of $70$~\cmv~pc on the total dispersion measure to the Large Magellanic Cloud (LMC). After subtracting the contribution from the Galactic disk they derive an upper limit ${\rm DM} \leq 23$~\cmv~pc on the dispersion measure due to hot thermal electrons along the line of sight to the LMC. 

%
%

\section{Corona Size and Mass}
\label{sec_motivation}

In this section we examine the basic properties of galactic coronae, as traced by observations of hot gas in the MW. What is the size of the Galactic corona? How massive is it? Can it contain the Galactic missing baryons? We show that a massive extended corona may be indicated by existing OVII observations. This is supported by the extended OVI distributions in the COS-Halos galaxies. The analysis presented in this section is preliminary to our full model we present in \S \ref{sec_model} and \S \ref{sec_results}.

\subsection{Hot Gas - Local or Extended?}
\label{subsec_mot_extent}

What is the spatial extent of the MW hot gas? \citet{Yao07} and \cite{Hagihara10} suggest that the OVII-bearing gas has a small path length of only a few kpc, and argue that the existing observations are consistent with absorption occurring entirely in the Galactic disk. \citet{Gupta12} reach the opposite conclusion that the gas path length is large, on the order of 100 kpc, and is evidence for an extended hot corona. The X-ray  absorption strength data alone cannot distinguish between the two scenarios since both the gas density and the path length are unknown. In principle, a combined analysis of absorption and emission should enable us to estimate these important quantities. However, as we now show, the assumptions regarding the hot gas metallicity and the existing OVII data can lead to very different conclusions.

The emission measure is given by 
\begin{equation}\label{eq:em}
EM \equiv \int{n_{\rm e} \nh dl} \equiv \ca \left< n_{\rm e} \right> \left< \nh \right>L~~~,
\end{equation}
and the hydrogen column density is
\begin{equation}\label{eq:N}
N_{\rm H}=\int{\nh dl} \equiv \cb \left< \nh \right> L~~~,
\end{equation}
where $\nh$ and $n_{\rm e}$ are the hydrogen and electron density, respectively, $\left<\nh \right>$ and $\left< n_{\rm e} \right>$ are their volume-averaged values and $L$ is the gas path length. The constants \ca~and \cb~depend on the radial dependence of the gas densities, and for uniform density gas, $\ca=\cb=1$. For our fiducial model we present in \S~\ref{sec_results}, for which there is a significant density gradient, $\ca=10.6$ and $\cb=2.3$.}  $EM$ is proportional to the observed emission line strength and $N_{\rm H}$ is proportional to the absorption line strength, and together can be solved for $n_{\rm H}$ and $L$. However, the translation of the measured quantities to $N_{\rm H}$ and $EM$ depends on the gas metallicity, $Z'$, the assumed solar oxygen abundance, $\ao$, and the OVII ion fraction, $f_{\rm OVII}$. For an observed OVII column density, the hydrogen column scales as $N_{\rm H} = N_{\rm OVII}/Z'$, and the emission measure scales as $EM = EM'/Z'$, where $EM'$ is the emission measure inferred from the observations assuming solar metallicity, $Z'=1$.

The hot gas path length can then be written as
\begin{multline}\label{eq:pathl}
L = 20.4 ~\kpc \left( \frac{\ca}{\chi_{\rm N}^2} \right) \left(\frac{N_{\rm OVII}}{10^{16}~\cmc}\right)^2 \left( \frac{f_{\rm OVII}}{0.5} \right)^{-2} \\
\times \left(\frac{EM'}{10^{-2}~{\rm cm^{-6}~pc}}\right)^{-1} Z'^{-1} ~~~.
\end{multline}
In this expression we have used the updated oxygen abundance, $a_O = 4.9 \times 10^{-4}$ \citep{Asplund09} and an OVII fraction, $f_{\rm OVII}$, normalized to 0.5. We adopt this value here as the OVII fraction since this is the value reached at temperature of $\sim 2 \times 10^6$~K, at which the OVIII fraction is non-negligble, as suggested by observations (see \S \ref{subsec_obs_ovii}). We stress that in our model (\S \ref{sec_model} and \S \ref{sec_results}) the ion fractions are not set to constant values, but are calculated assuming CIE for each gas temperature. 

In their analysis, \citet{Hagihara10} present measurements of the OVII column density ($N_{\rm OVII}=5.9 \times 10^{15}$~\cmc, as measured from the spectrum of PKS 2155-304) and an emission measure of $7.5 \times 10^{-3}~{\rm cm^{-6}~pc}$ (after updating the solar oxygen abundance, originally $EM = 4.35 \times 10^{-3}~{\rm cm^{-6}~pc}$). Inserting these values into Eq.~\ref{eq:pathl} and assuming solar metallicity and uniform density gas, we get $L = 9.4$~kpc, consistent with metal-enriched gas in the vicinity of the disk.

However, when examining this result two questions need to be addressed. The first is whether the observed values used by \citet{Hagihara10}, based on a single sightline, are typical. The PKS-2155 OVII column density is low compared to the typical value we adopt in \S \ref{sec_obs}, $N_{\rm OVII}=1.4 \times 10^{16}$~\cmc, based on a sample of $43$ sightlines by \citet{Fang15}. For the EM, \citet{Henley10} model the X-ray spectrum observed in 26 fields and infer the emission measure of the hot gas. In their model they assume a single temperature gas and a solar metallicity.  We correct the measured values reported by H10 to the updated oxygen abundance, $\ao = 4.9 \times 10^{-4}$, increasing them by a factor of $\sim 1.74$. The median $EM'$ of the sample (and $90\%$ confidence error range), as estimated by H10 and after the abundance correction, is $3.3~(2.5-4.0) \times 10^{-3}~{\rm cm^{-6}~pc}$, lower than inferred for PKS-2155 by \citet{Hagihara10}. \cite{Yoshino09} report measurements similar to H10, for a smaller sample of 12 fields. Inserting our adopted value for the column density and the H10 emission measure into Eq.~\ref{eq:pathl} results in a significantly larger path length of $L \sim 120$~kpc.

Second, we can ask how this result changes for different assumptions on the gas metallicity. Reducing the metallicity, as possibly appropriate for CGM environments (\citealp{Tripp03,Fox05,Lehner13,Suresh15}) leads to a larger path length. For example, $Z' = 0.5$, the preferred value we find for our fiducial model (see \S \ref{sec_results}), increases the path length by a factor of 2, to $240$~kpc, comparable to the estimated virial radius of the MW.

These considerations show that an extended origin is more likely for much of the hot OVII. As we discuss in \S\ref{sec_model} and \S \ref{sec_results}, in our picture the OVII is tracing an extended corona.

\subsection{Mass \& Baryon Fraction}
\label{subsec_mot_fraction}

With the aid of Eq.~\ref{eq:pathl}, the mean hot gas density along the path length, $\left< n_{\rm H} \right> = N_{\rm H}/\cb L$, can be written as
\begin{multline}
\left< n_{\rm H} \right> = 7.56 \times 10^{-4} \left( \frac{\cb}{\ca} \right) \left( \frac{N_{\rm OVII}}{10^{16}~\cmc} \right)^{-1} \left( \frac{f_{\rm OVII}}{0.5} \right) \\
\times \left( \frac{EM'}{10^{-2}~{\rm cm^{-6}~pc}} \right) ~~\cmv ~~~.
\end{multline}
Given the gas density and path length we can estimate the total hot gas mass traced by OVII. 
For a corona extending to radius $L$, the total gas mass is 
\begin{multline}\label{eq:mcor}
M_{\rm tot} = 5.7 \times 10^8 \left( \frac{\chi_{\rm EM}^2}{\chi_{\rm N}^5} \right) \left( \frac{{\widetilde m}}{m_{\rm H}} \right)
\left( \frac{N_{\rm OVII}}{10^{16}~\cmc}  \right)^5 \left( \frac{f_{\rm OVII}}{0.5} \right)^{-5} \\
\times \left( \frac{EM'}{10^{-2}~{\rm cm^{-6}~pc}} \right)^{-2} Z'^{-3} ~~ \msun ~~~ ,
\end{multline}
where $\widetilde{m}$ is the mass per hydrogen nucleus. For our typical values of $EM$ and $N_{\rm OVII}$, with $Z'=0.5$ and $\tilde{m}=4/3~m_{\rm H}$~(for a primordial helium abundance) and a uniform density corona ($\cb=1$, $\ca=1$), the resulting total hot gas mass is $3.0 \times 10^{11}$ \msun. The mass derived here is an upper limit on the total gas mass, since if the hot gas has a non-uniform density, the emission is dominated by the denser regions.

To address the baryon budget of the Milky Way, we need to know its total mass. \cite{BK13} constrain the MW virial mass to be between $1.0$ and $2.4 \times 10^{12}$ at a confidence level of $90\%$ (see also \citealt{Pen16} and \citealt{Li08} for estimates at the low and high limits of this range). \cite{Sakamoto03} note that the upper mass limit is strongly affected by the dwarf galaxy Leo~I, increasing it from $1.8$ to $2.5 \times 10^{12}$~\msun. For our missing baryons estimates we adopt a range of $1-2 \times 10^{12}$~\msun.

Given the total mass of the Galaxy, we assume it is the sum of the dark matter and the baryonic mass. 
According to the \cite{Planck16}, the cosmological baryon fraction, $\fb=0.157$. 
For this simplified analysis we assume that the observed baryonic mass, $M_{\rm b,obs}$, of the Galaxy is 
$\sim 6 \times 10^{10}$~\msun, dominated by the material in the disk, including stars, gas and dust (see \citealp{Draine11}). This value agrees with other estimates of the observed baryonic mass, with $M_{\rm b,obs} = 5.9 \pm 0.5 \pm 1.5 \times 10^{10}$~\msun~(statistical and systematic errors, respectively) by \cite{Bovy13} and $M_{\rm b,obs} = 6.08 \pm 1.14 \times 10^{10}$~\msun, by \cite{Lic15}. We can then write the missing baryon mass as $M_{\rm b,miss} = \fb\mvir - M_{\rm b,obs}$. For our adopted virial mass range, $M_{\rm b,miss} = 1.0-2.5 \times 10^{11}$~\msun. The upper limit on the hot gas mass we estimate using Eq.~\ref{eq:mcor} is above this range and we conclude that an extended corona can account for a significant fraction of, if not all the missing baryons in the MW. Motivated by this analysis, in the next sections we construct our model for the corona and present results for a specific set of fiducial parameters.

%
%

\section{Model: Warm/Hot Corona}
\label{sec_model}

We construct our model corona based on simple physical assumptions regarding the gas properties and guided by the results of hydrodynamics simulations.

The main ideas are as follows. The coronal gas is multi-phased, with two components - hot and warm. The hot gas is the origin of the OVII and OVIII absorption and X-ray emission, and the warm component is the origin of the OVI absorption. Both components have log-normal distributions of temperatures and densities, representing turbulence and motivated by theoretical work results of hydrodynamics simulations (see \S\ref{subsec_model_dist}). For the hot gas, the cooling efficiency is low with a long cooling time ($\gtrsim 10^9$~years) that allows the formation of a quasi-static corona around the galaxy (see \S\ref{subsec_model_hse}). The warm component is gas that cooled out of the higher-density tail of the turbulent hot component (see \S\ref{subsec_model_cool}). The cooling time of the warm gas is short and requires a heat source to maintain the mass of OVI absorbing gas (see also \S\ref{subsec_disc_mass}).

The gas metallicity is an important parameter, affecting both the gas cooling rates (see \S\ref{subsec_model_cool} and see \S\ref{subsec_model_spectrum}), gas columns, emission measures (\S\ref{subsec_model_ocol})  and masses. We assume a spatially constant metallicity throughout and consider values in the range of $Z'=0.1-1.0$.

\subsection{Hydrostatic Equilibrium}
\label{subsec_model_hse}

We consider a corona of hot gas in hydrostatic equilibrium (HSE) in the gravitational potential of a galaxy-size dark matter halo (see also \citealp{Tepper15}). We neglect the self-gravity of the coronal gas since it makes a relatively small contribution to the total mass (verified for self-consistency). For the model we present here we adopt the enclosed mass distribution, stellar disk plus NFW dark matter halo, favored by \cite{Klypin02} (hereafter K02) for the Galaxy. K02 consider~\mvir~in the range of $0.7-2.0 \times 10^{12}$~\msun, all consistent with observations (see also \S \ref{subsec_mot_fraction}). In the K02 favored model we adopt in this work, $\mvir = 10^{12}$~\msun~and the virial radius is $\sim 250$~kpc.  The maximum circular velocity is $223$~\kms. We assume spherical symmetry from 8 to 250 kpc and neglect the non-spherical shape of the potential close to the Galactic disk.

In our model the hot gas distribution extends out to large radii, as suggested by the OVI distribution in the COS-Halos galaxies.
\cite{More15} use dark-matter-only simulations of cluster- and group-sized halos and find that the dark matter profile steepens significantly at $\sim\rvir$ (see also \citealp{Patej16}). \cite{Lau13p1} find that the observed gas profiles of clusters steepen at the same radius as the dark matter profiles. We define the outer radius of the corona, \rcgm, and for the purpose of this work, set it to be the virial radius,  \rcgm=\rvir. Simulations show that the density decreases beyond this radius, as well as the gas temperature, before reaching the virial shock \citep{Birnboim03,Genel14}. Hence, we assume that the contribution from gas at these radii, $r>\rvir$, to the total mass of the corona and to the absorption and emission along a given line of sight, will be negligible.

There is some uncertainty regarding the virial radius of the Milky Way ($250^{+60}_{-30}$~kpc, see \citealp{Busha11}) and the actual size of the corona may be slightly larger than \rvir~(by $\sim 10-20\%$), as suggested by observations (\citealp{Lehner15,Tully15}) and simulations including  feedback processes (\citealp{Sokol16,WDS16}). We address the affect of \rvir~and the corona size on the model results in our next work (paper II).

We assume that the mean gas temperature, $\left< T \right>$, is constant as a function of radius. The thermal velocity dispersion in the gas is then given by $\sigma_{\rm th} = \sqrt{k_B \left<T \right>/\bar{m}}$, where $\bar{m}$ is the mean particle mass and $\kb$ is Boltzmann's constant. As we show in \S \ref{subsec_model_dist}, the temperature setting $\sigma_{\rm th}$ is the mass-weighted mean temperature. Motivated by the distinct OVI velocities between different line components (\S \ref{subsec_obs_ovi}) we also add a turbulent component to the velocity dispersion (see \S \ref{subsec_obs_ovi}). Finally, we allow for additional support by cosmic rays (CR) and/or magnetic fields. 
We then write 
\begin{equation}\label{eq:sigma}
P/\rho = \alpha \sigma_{\rm th}^2 + \sigma_{\rm turb}^2
\end{equation}
relating the gas pressure $P$, the gas mass density $\rho$, and the thermal and non-thermal velocity dispersions. The factor $\alpha$ accounts for the possible CR and magnetic support. Neglecting self-gravity for the gas, the equation of hydrostatic equilibrium (HSE) is then 
\begin{equation}\label{eq:hse}
\frac{dP}{P} = -\frac{d\varphi}{\alpha \sigma_{\rm th}^2 + \sigma_{\rm turb}^2} ~~~,
\end{equation}
where $\varphi$ is the gravitational potential, dominated by the dark matter halo and the central disk.
Integration gives our hot gas pressure profile for the corona
\begin{equation}\label{eq:pprof}
P(r) = P_0 \exp\left[-\int_{r_0}^r{\frac{G}{\alpha \sigma_{\rm th}^2 + \sigma_{\rm turb}^2}\frac{M(r')}{r'^2}dr'}\right] ~~~,
\end{equation}
where $P_0$ is a normalization pressure at some inner radius $r_0$. Here $P_0$ is the total pressure (thermal, turbulent and other), given by 
\begin{equation}
P_0 = \alpha P_{0,\rm th} + P_{0, \rm turb} = P_{0,\rm th} \left( \alpha + \frac{\sigma_{\rm turb}^2}{\sigma_{\rm th}^2} \right) ~~~,
\end{equation}
where $P_{0, {\rm th}}$ and $P_{0, {\rm turb}}$ are the thermal and turbulent pressures at $r_0$, respectively.

Since our model does not include a gaseous disk component, we are interested in the pressure outside the disk for the pressure normalization, $P_{0,\rm th}$. \cite{Dedes10} use 21~cm observations of HI clouds above the disk (at $1<z<5$~kpc, and at distances of $10<d<15$~kpc from the Galactic center) to estimate the pressure around them at $P/\kb \sim 500-1500~\cmv$~K. 
This is lower than the standard value in the Galactic plane, $P_{0,\rm th}/\kb = 3000$~K~\cmv, estimated by \cite{Wolfire03} at the solar circle, $r_0=8.5$~\kpc. In this work we adopt a value of $P_{0,{\rm th}}/\kb =2200~\cmv$~K at $r_0=8.5$~\kpc. We examine the effect of the assumed normalization pressure on the model results in our paper II.

For the turbulent velocity scale, we choose a value for $\sigma_{\rm turb}$ within the range of $60-80$~\kms, as estimated from the OVI line velocities (see \S\ref{subsec_obs_ovi}). For $\alpha$ we select a value between $1$ and $3$. For $\alpha=1$ there is only thermal and turbulent support, while assuming equipartition between the thermal, cosmic rays and magnetic energy gives $\alpha=3$. \cite{OML10} and \cite{KKO11} include the turbulence in this factor so that $\alpha_{\rm OML} \equiv \alpha + \sigma^2_{\rm turb}/\sigma_{\rm th}^2$. They estimate $\alpha_{\rm OML} \approx 5$ in the Galactic disk.

\subsection{Local Log-Normal Distribution}
\label{subsec_model_dist}

We assume that the mean temperature is constant on the large scale of the corona but that fluctuations in temperature and density arise on small scales. The observed wide range of oxygen ions suggests empirical evidence for temperature variations in the CGM. Density and temperature fluctuations on small scales are found by \citet{McCourt12} and \citet{Konstandin16} in hydrodynamics simulations.

We allow for a range of temperatures in the gas, assuming a (volume-weighted) log-normal distribution
\begin{equation}\label{eq:Tdist}
g_{\mbox{\tiny {\it V}}}(x) dx = \frac{1}{s \sqrt{2\pi}}e^{-x^2/2s^2} dx ~~~.
\end{equation}
Here $x\equiv \ln{(T/T_{\rm med, {\mbox{\tiny {\it V}}}})}$, where $T_{\rm med, {\mbox{\tiny {\it V}}}}$ is the median of the distribution and $s$ is the width. 
As functions of temperature the distribution functions are then defined by $p(T)dT \equiv g(x)dx$ (both for the volume- and mass-weighted cases). For the hot gas, we consider $T_{{\rm med}, {\mbox{\tiny {\it V}}}}$ in the range $1-3 \times 10^6$~K, around the virial temperature of the MW halo. \cite{Blais93} and \cite{Pad97} provide theoretical motivation for log-normal density distributions in turbulent media.

We assume isobaric fluctuations in temperature in the hot gas. This implies that the density fluctuations are also log-normal.
The local pressure, $P$, is related to the local mean density and temperature through the relation
\begin{equation}\label{eq:Plocal}
P \equiv \left< \rho T \right>_{{\mbox{\tiny {\it V}}}} = \left<\rho\right>\left< T \right>_{{\mbox{\tiny {\it M}}}} ~~~,
\end{equation}
where $\left<\rho\right>$ is the (volume-weighted) mean density, and $\left<T\right>_{\mbox{\tiny {\it M}}} = T_{\rm med,\mbox{\tiny {\it V}}} \times e^{-s^2/2}$ is the mass-weighted mean temperature (see Appendix). Thus, it is the mass-weighted mean $\left<T\right>_{\mbox{\tiny {\it M}}}$ that determines $\sigma_{\rm th}$ in the equation of hydrostatic equilibrium (Eq.~\ref{eq:hse}).

\subsection{Warm and Hot Components}
\label{subsec_model_cool}

The hot gas cools radiatively, with the emission rate per unit volume given by $\mathcal{L} = \Lambda(T) n_{\rm e} n_{\rm H}~({\rm erg~cm^{-3}~s^{-1}})$, where $\Lambda (T)$ is the radiative cooling efficiency and $n_{\rm e}$ and $n_{\rm H}$ are the electron and hydrogen densities, respectively. For a given gas metallicity, $\Lambda(T)$ is a function of temperature only and we use the cooling functions from \citet{GS07}. The cooling time depends on the temperature and density of a specific ``cell'' in the log-normal distribution and for a primordial helium abundance can be written as (see Eq.~8 in \citealp{GS07})
\begin{equation}\label{eq:tcool}
\tcool = \frac{4.34 \times (3/2+\delta) \kb T }{n \Lambda(T)} ~~~.
\end{equation}
Here $n$ is the total number density of particles and $\delta$ can have a value of $0$ or $1$ for isochoric or isobaric cooling, respectively. We can compare \tcool~to the local dynamical time at any $r$, $\tdyn \simeq \sqrt{r^3 / GM(r)}$. If the typical cooling time is longer than \tdyn, the gas is expected to be in hydrostatic equilibrium and the cooling is isobaric.

\begin{figure}
\includegraphics[width=0.45\textwidth]{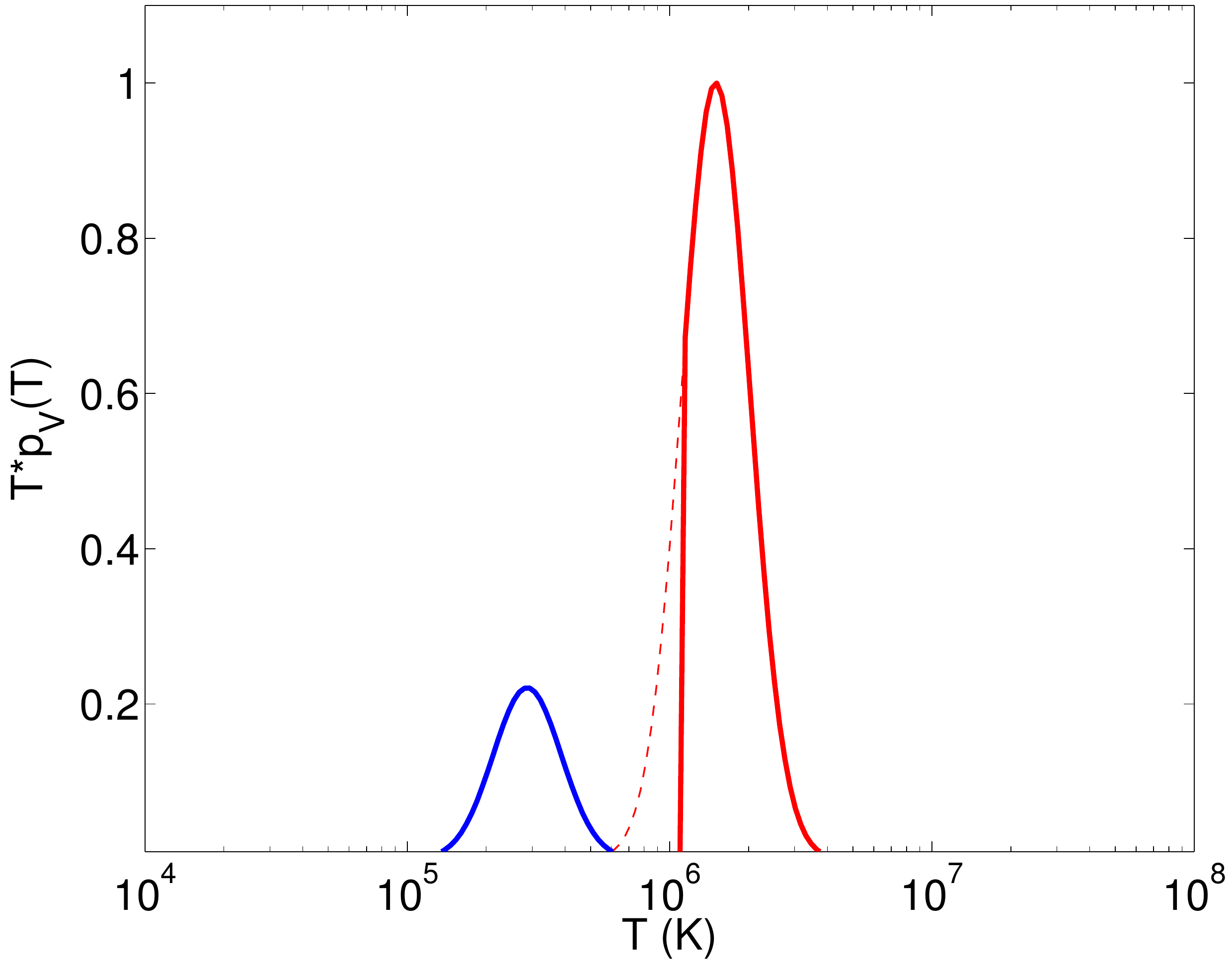}
\caption{The gas probability distribution as a function of temperature. The function shown here is $T\ppv(T) = \gv(x)$ (see Eq.~\ref{eq:Tdist} and the text below it). The hot component (red) distribution has a minimal temperature limit ($T_{\rm min}$, see \S \ref{subsec_model_cool}). At low temperatures (thus high densities, for isobaric fluctuations) the gas cools rapidly and forms the warm component (blue).  The dashed red line shows the full log-normal distribution of the hot component before truncation at $T_{\rm min}$. The minimal temperature is a function of radius and the value displayed here is for $r=200$~kpc. The OVII and OVIII ions are formed in the hot component, and OVI - in the warm gas.}
   \label{fig:prob}
 \end{figure}

For some cells, however, \tcool~ may be shorter than \tdyn~and the gas can cool. We therefore calculate the ratio $\tcool / \tdyn$. Summing over gas cells for which this ratio falls below a defined value, an input parameter in our model, gives the gas mass which cools to form the lower temperature warm component. This occurs for gas cells below some temperature, $\Tmin$, the minimum temperature of the hot gas distribution. The reason for this is that cells at lower temperatures have higher densities (since the hot gas is isobaric) and higher cooling efficiencies (in the relevant temperature range, around $10^6$~K). \citet{GS07} show that for hot gas cooling radiatively, the cooling is isobaric at first but eventually becomes isochoric, for a wide range of gas densities and cloud sizes (see their Table~2 and Figure~1). Thus, we assume that the warm component is formed isochorically in the sense that it occupies the same volume as the hot gas with $T<\Tmin$ from which it cooled. The volume filling factor of the warm gas is then 
\begin{equation}\label{eq:fill_fac}
\fvw(r)=\int_0^{\Tmin} \ppv(T) \mathrm{d}T ~~~,
\end{equation}
where $\ppv(T)$ is the hot gas volume-weighted probability distribution. The fraction of the mass that is warm, $\fmw$, is given by a similar expression, with $\ppv(T)$ replaced by $\ppm(T)$, the mass-weighted distribution (see Appendix).

We assume further that the processes that maintain the temperature of the warm component also result in a log-normal distribution of temperature and density and that during the formation of the warm component, the transition from isobaric to isochoric cooling is immediate and simultaneous for all cells. Thus, the warm gas cells are at a fixed pressure, $P_{w}$, that is less than that of the ambient hot gas by a factor $n_f T_f/n_i T_i = T_f/T_i = {\rm constant}$, where the subscripts $i$ and $f$ stand for the densities and temperatures before and after the cooling, respectively. In reality, the warm gas may have a range of reduced pressures. However, the pressure affects only the cooling luminosity of the warm gas, which is currently unobservable (see \S \ref{subsec_results_mass} and Figure \ref{fig:spectrum}), but not the observable OVI column density. 

Finally, we calculate the total cooling times of the hot and warm components as the ratios of the total thermal energies to the total luminosities to estimate the lifetimes of these coronal structures. As described earlier, the hot component cools isobarically. The transition from the hot to the warm component is isochoric and the warm component cools isochorically. For a fully isobaric transition from hot to warm gas the luminosity is increased by the ratio $(n_f/n_i)^2 = (T_i/T_f)^2$.

The warm/hot gas mixing we are invoking is motivated by the observed flat OVI column density profile (see Figure \ref{fig:ovi}). Importantly, a hydrostatic distribution of warm gas with $T \sim 3 \times 10^5$~K (where the OVI ion fraction peaks) in an MW-sized  dark matter halo would form a much steeper density profile, inconsistent with observations. Our model provides a natural explanation for the extended OVI.

As shown in Figure \ref{fig:prob}, the temperature distribution of the hot gas is a truncated log-normal and as a result, the mean and median temperatures of the hot gas are higher than the mean and median temperatures of the underlying log-normal. This is a small effect, however, since only a small fraction of the hot gas has cooled (see \S \ref{sec_results}).

\subsection{Oxygen Column Densities}
\label{subsec_model_ocol}

The model is set up once we compute the distributions of gas temperatures and densities as a function of radius. To find the oxygen column densities for a given (constant) metallicity, we adopt the solar relative oxygen abundance from \citet{Asplund09}, ${\rm [O/H]} = 8.69$ ($\ao=4.9 \times 10^{-4}$). We then calculate the oxygen column densities, for comparison with the observations presented in \S \ref{sec_obs}.

For the oxygen ion fractions we assume the gas is in collisional ionization equilibrium (CIE) and we adopt the fractions calculated by \cite{GS07} as a function of temperature. In \S \ref{subsec_disc_photo} we show that in our model, photoionization (by the metagalactic radiation) does not play a significant role even for the OVI. We verify our assumption of CIE by calculating the ratio of the cooling time, $\tcool$, given in Eq.~\ref{eq:tcool}, to the recombination time, $\trec$, given by $1/n \alpha^{\rm tot}_i$, where $\alpha^{\rm tot}_i$ is the recombination rate coefficient for a given ion. Non-equilibrium effects become significant if $\tcool/\trec<1$. Using the recombination rate coefficients from \cite{Nahar03}, we find that $\tcool/\trec \sim 100$ and $4$ for the hot and warm components, respectively. Thus, our assumption of CIE is valid for the hot gas and reasonable for the warm gas.

For OVII and OVIII, we calculate the column density from the Sun's location in the Galaxy (at $r_0=8.5$~kpc from the center) to the outer limit of the corona along directions that span the observed sightlines. For example, for the hot component, the OVII column density is given by
\begin{multline}\label{eq:ovii_col}
N_{\rm OVII} = \ao Z' \\
\times \int_{r_{0}}^{\rcgm}{\mathrm{d}l}\int_{\Tmin}^{\infty}{\ppv(T) n_{\rm H}(r,T)f_{\rm OVII}(T) \, \mathrm{d}T} ~~~,
\end{multline}
where $dl$ is the (angular dependent) path element along the sightline, $\ao$ is the solar oxygen abundance relative to hydrogen, $f_{\rm OVII}$ is the OVII ion fraction and $\ppv(T)$ is the temperature probability distribution. The contribution from the warm component is calculated similarly, but it is only a small fraction of the total OVII column. The gas density, $n_{\rm H}(T) \propto T^{-1}$, with our assumption of local isobaric fluctuations. The expression for OVIII is identical but with the respective ion fraction, $f_{\rm OVIII}(T)$ in the integrand.

The OVI sightlines are through the corona, as viewed by an external observer. 
The column density at impact parameter $h$ is given by 
\begin{multline}\label{eq:ovi_col}
N_{\rm OVI}(h) = 2 \ao Z'  \\
\times \int_{0}^{\rcgm} \fvw(r) {\mathrm{d}l}\int_{0}^{\infty}{\pvw(T) n_{\rm H}(r,T) f_{\rm OVI}(T)\, \mathrm{d}T} ~~~,
\end{multline}
where $dl = r \,\mathrm{d}r / \sqrt{r^2-h^2}$ is the path element along the line of sight, $f_{\rm OVI}$ is the OVI ion fraction and $\pvw(T)$ and $\fvw(r)$ are the temperature probability distribution and the volume filling factor of the warm component, respectively (see Eq.~\ref{eq:fill_fac}). We have assumed that the turbulence and reheating processes in the warm gas can heat it above $\Tmin$, so the integral over the temperature distribution extends to $\infty$.

The dispersion and emission measures are calculated similarly. 
We calculate the dispersion measure to the LMC, at $d_{\rm LMC} \approx 50$~kpc \citep{Freedman12}, for comparison with the limit derived by \cite{Anderson10} (see \S \ref{subsec_obs_other}). The dispersion measure is given by
\begin{equation}\label{eq:DM}
DM_{\rm LMC}= \int_{r_{0}}^{d_{\rm LMC}} {\mathrm{d}l}
\int_{0}^{\infty}{\ppv(T) n_{\rm e}(r,T) \, \mathrm{d}T} ~~~,
\end{equation}
where $dl$ is the path element along the sightline from the solar circle to the LMC and $n_{\rm e}$ is the total electron density (of the hot and warm gas). 

The emission measure is given by
\begin{equation}\label{eq:EM}
EM = \int_{r_{0}}^{\rcgm} {\mathrm{d}l}
\int_{T_{\rm min}}^{\infty}{\ppv(T) n_{\rm hot}(r,T)n_{\rm e}(r,T) \, \mathrm{d}T} ~~~.
\end{equation}
Since the observed EM values are inferred by modeling spectra in the X-ray, here $n_{\rm H}$ and $n_{\rm e}$ are the hydrogen and electron densities of the hot component (with $T_{\rm min}$ as the lower integration limit), and we integrate out to \rcgm=\rvir.

\subsection{Emission Spectrum}
\label{subsec_model_spectrum}

We calculate the spectrum of the corona in emission for comparison with observations \citep{Strickland04a,Tullman06a}.
To calculate the emission spectrum of each component, we start with the emissivity of a gas ``parcel" at a given density $n$, temperature $T$ and metallicity $Z'$. We integrate over the temperature probability distribution in our model to find the spectrum emitted at each radius. Integrating this over the volume results in the total spectrum of the hot corona, given by
\begin{equation}\label{eq:spec1}
J(\nu) = 4\pi \int_0^{\rcgm}{r^2 \mathrm{d}r} \, \int_{T_{\rm min}}^{\infty}{j_{\nu}(n,T) \ppv(T) \, \mathrm{d}T} ~~~,
\end{equation}
where $j_{\nu}$ is the gas emissivity, $\ppv(T)$ is the temperature probability distribution (defined below Eq.~\ref{eq:Tdist}) and $J(\nu)$ has units of ${\rm erg~s^{-1}~Hz^{-1}~sr^{-1}}$.
Since the gas density in the model is low ($<10^{-2}$~\cmv), we use the coronal approximation for the variation of emission intensity with gas density. We calculate the spectrum of gas at a given density, $n_0$, and scale the emission intensity with the density squared to obtain the spectra for the different cells in the distribution. The spectrum of a gas parcel with density $n$ and temperature $T$ is thus given by 
\begin{equation}\label{eq:spec2}
j_{\nu}(n,T) = \frac{n^2(T)}{n^2_0} j_{\nu}(n_0,T) ~~~,
\end{equation}
where $n_0 = 10^{-4}$~\cmv. 
We used version 13.00 of Cloudy, last described by \citep{Ferland13}, to calculate a set of (optically thin) emissivity spectra, $j_{\nu}(n_0,T)$, assuming CIE (see \S\ref{subsec_model_ocol}), for the range of temperatures in our model. The resulting spectrum allows to compare our model with observations of galactic coronae in emission in the 0.3-2.0 keV band.

Next we calculate the emission intensity in the \lia~and \lib~oxygen features, for comparison with the observations presented in \S\ref{subsub_lines}. First, we use Cloudy to compute the line emissivities for the transitions summarized in Table~\ref{tab:xray_atomic} as a function of temperature, $j(T)$. For each transition we then integrate over the gas density distribution and along sightlines from the Sun's location to the corona outer limit (similar to the OVII and OVIII column densities calculation). The OVII $21.60~\AA$ transition line intensity, for example, is then given by
\begin{multline}\label{eq:li1}
I_{21.60~\AA} = 
\ao Z'  \int_{r_{0}}^{\rcgm} {\mathrm{d}l} \int_{0}^{\infty}{\frac{j(T)}{h\nu} \ppv(T) f_{\rm OVII}(T)\, \mathrm{d}T} ~~,
\end{multline}
where $j(T)$ is the line emissivity per gas particle, and the line intensity is in units of $ \rm photons~cm^{-2}~s^{-1}~sr^{-1}$. Most of the relevant transitions presented in Table~\ref{tab:xray_atomic} belong to the OVII ion, and for them the expression is identical with the respective emissivities. In the expression for the OVIII $18.97~\AA$ line, contributing to the \lib~feature, we use the OVIII ion fraction instead of the OVII. In our calculations, we use the CIE temperature-dependent ion fractions calculated from \cite{GS07}. Finally, we sum over the different transitions contributing to each feature.

We present our fiducial model for the corona in the next section and compare it with X-ray absorption and emission observations in \S~\ref{subsec_results_obs}.

%
%

\section{Fiducial Model}
\label{sec_results}

 \begin{figure}
\includegraphics[width=0.45\textwidth]{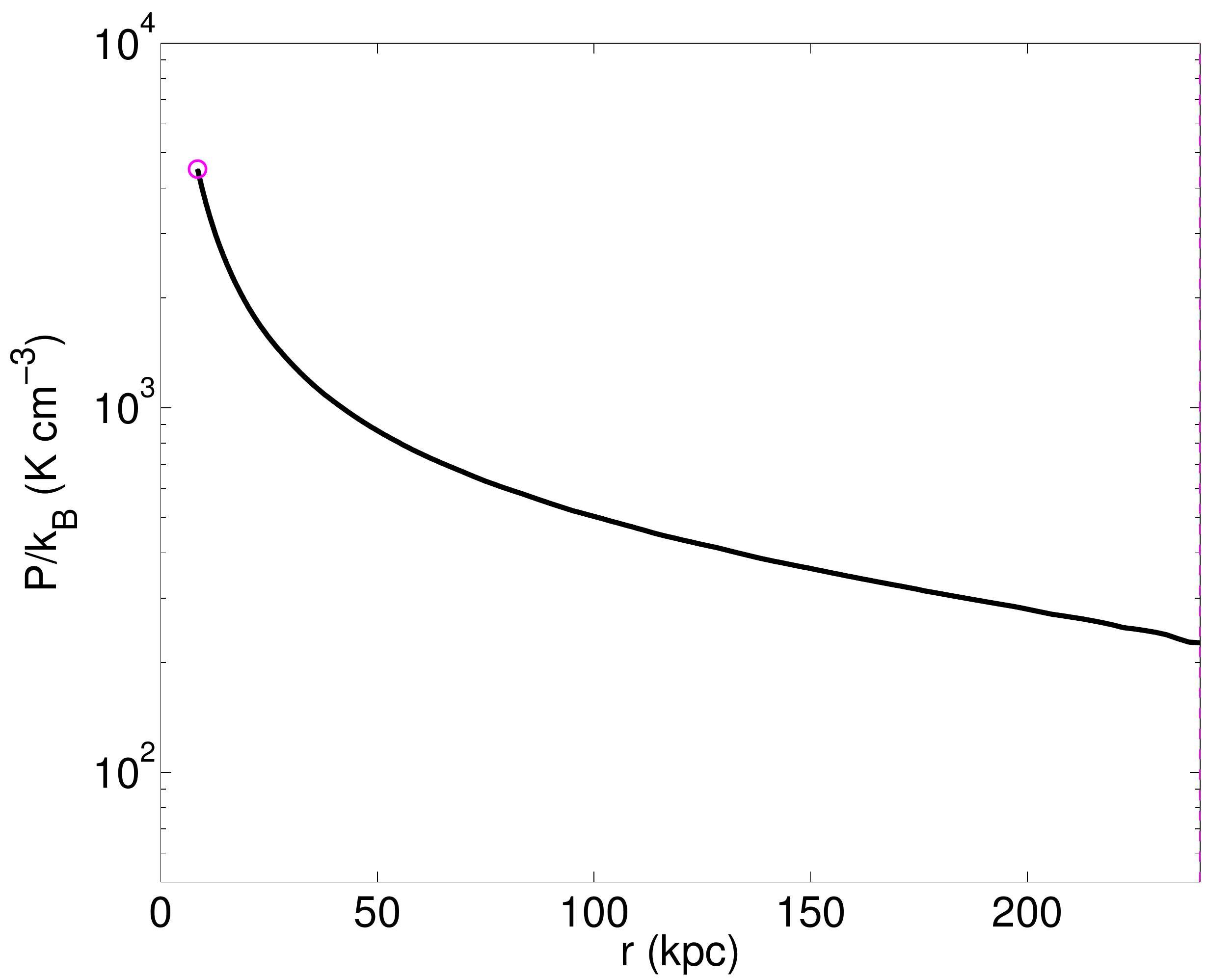}
\caption{Coronal gas pressure profile, dominated by the hot component. The solar circle ($r_0=8.5$~kpc, magenta circle) is marked here and in the next figures presenting corona profiles. The thermal (total) pressure at $r_0$ is normalized to $P_{\rm 0,th}/\kb = 2200$ ($4580$)~\cmv~K, and the hot gas pressure at the virial radius is $\sim 230 $~\cmv~K (see \S \ref{subsec_results_dist} for details).}
   \label{fig:pressure}
 \end{figure}

In this section we present our results for a single fiducial model that provides a good match to the observations. The main input parameters and resulting properties of the model are listed in Table~\ref{tab:mod_par}. For example, for our fiducial model, the volume-weighted median temperatures (see Eq.~\ref{eq:Tdist}) for the hot and warm components are equal to $1.5 \times 10^6$ and $3.0 \times 10^5$~K, respectively. The corresponding mass-weighted median temperatures are $1.4 \times 10^6$ and $2.7 \times 10^5$~K. The gas metallicity in our model is $Z'=0.5$.

Table~\ref{tab:mod_res} presents the main observable quantities of the CGM, comparing the measured values (discussed in \S \ref{sec_obs}) and the results of our fiducial model, presented in this section.

\bgroup
\def\arraystretch{1.5}

\begin{table*}
\centering
	\caption{Fiducial model - Summary of properties}
	\label{tab:mod_par}

    \begin{tabular}[t]{c c}

		\begin{tabular}{| l | c c|}
			\midrule
			\multicolumn{3}{|c|}{Input Parameters}			\\
			 \midrule
 			$\mvir$ 									& 		\multicolumn{2}{c |}{$10^{12}~\msun$}								\\
			$\rvir$ 									& 		\multicolumn{2}{c |}{$250~\kpc$}										\\
 			$\rcgm$ 									& 		\multicolumn{2}{c |}{$250~\kpc ~(=\rvir)$}						\\
			Metallicity									& 		\multicolumn{2}{c |}{$Z' = 0.5$}											\\
			$\sigma_{\rm turb}$ 				& 		\multicolumn{2}{c |}{$60$~\kms}										\\
			$\alpha$ ($\alpha_{\rm OML}$)		
			\footnote{~$\alpha$ is the ratio of cosmic ray and magnetic filed pressure to the thermal presure.
			$\alpha_{\rm OML}$ also includes the turbulent pressure (see \S \ref{subsec_model_hse}).}
															& 		\multicolumn{2}{c |}{$1.9$ ($2.1)$}				\\
			$\tcool/\tdyn$ 				& 		\multicolumn{2}{c |}{$8$}										\\
			Pressure at $r_0$ 					& 		\multicolumn{2}{c |}{$P_{0,{\rm th}}=2200$~K~\cmv}		\\
															& 		\multicolumn{2}{c |}{$P_0\sim 4580$~K~\cmv}					\\
			\midrule
			 & Hot Gas & Warm Gas 	\\
			\midrule
			$T_{{\rm med},{\mbox{\tiny {\it V}}}}$ [K]	& $1.5 \times 10^6$ 	& $3.0 \times 10^5$ 	 \\
			$T_{{\rm med},{\mbox{\tiny {\it M}}}}$ [K]	& $1.4 \times 10^6$ 	& $2.7\times 10^5$ 	 \\
						$s$
			\footnote{~See Eq.~\ref{eq:Tdist} for the gas probability distribution and the definition of $s$.}
			 												& $0.3$							& $0.3$ 						 				\\
			\midrule
    \end{tabular} &
    
    \begin{tabular}{| l | c c|}
			\toprule    									
			\multicolumn{3}{|c|}{Results}			\\			
			\midrule
			$M_{\rm gas}(\rvir)$				& 		\multicolumn{2}{c |}{$1.2 \times 10^{11}~\msun$} 	\\
			$\left< \nh\right>_{\rvir}$		& 		\multicolumn{2}{c |}{$4.6 \times 10^{-5}~\cmv$} 		\\
			$\nh(50-100 \rm~{kpc})$				& 		\multicolumn{2}{c |}{$0.83-1.3 \times 10^{-4}$~\cmv}		\\
			$\left< \fvw \right>$				& 		\multicolumn{2}{c |}{$0.15$}					\\
			$P(\rvir)$									& 		\multicolumn{2}{c |}{$230$~K~\cmv}		\\
			$N_{\rm H}$								& 		\multicolumn{2}{c |}{$8.7 \times 10^{19}$~\cmc}			\\
			$EM$										& 		\multicolumn{2}{c |}{$6.4 \times 10^{-3}~{\rm cm^{-6}~pc}$}		\\
			$\ca$										& 		\multicolumn{2}{c |}{$10.6$}						\\
			$\cb$										& 		\multicolumn{2}{c |}{$2.3$}  					\\
			
			\midrule
			 & Hot Gas & Warm Gas 	\\
			\midrule
			$M_{\rm gas}~[\msun]$ 						& $9.4 \times 10^{10}$  & $2.7 \times 10^{10}$\\
			$L ~[\ergs]$ 										& $5.6 \times 10^{41}$  & $7.8 \times 10^{41}$\\
			$\tcool~[\rm years]$	  						& $6.8 \times 10^{9}$  & $2.2 \times 10^8$	 \\
			\bottomrule
		\end{tabular}\\
    
	\end{tabular}
\end{table*}

\begin{table*}
\centering
	\caption{Fiducial model - Comparison to observations}
	\label{tab:mod_res}
		\begin{tabular}{| l || c c|}
			\midrule
			& Observations (\S \ref{sec_obs}) & Fiducial model 	(\S \ref{sec_results}) \\
			 \midrule
			$N_{\rm OVII}~(\cmc)$						& 	$1.4~(1.0-2.0) \times 10^{16}$			& $1.6\times 10^{16}$			\\
			$N_{\rm OVIII}~(\cmc)$						& 	$0.36~(0.22-0.57) \times 10^{16}$	& $3.8\times 10^{15}$			\\
			$\rm OVII/OVIII$~ratio						& 	$4.0~(2.8-5.6) $									& $4.5$									\\							
			$\sigma_{\rm oxygen}$ (~\kms)			& 	67.2 (54.5 - 79.7)								& $72.0$									\\
			$DM$ (LMC)~(${\rm \cmv~pc})$			& 	$\lesssim 23$										& $17.4$									\\
			$S_{0.4-2.0}~(~\fluxun)$						& 	$2.1~(1.9-2.4) \times 10^{-12}$		& $0.82 \times 10^{-12}$		\\
			$22~\rm \AA~(\liun)$							& 	$2.8~(2.3-3.4) $									& $1.2$									\\
			$19~\rm \AA~(\liun)$ 						& 	$0.69~(0.58-0.83) $							& $0.33$									\\
			$\lia/\lib$~ratio									& 	$4.3~(3.4-5.5) $									& $3.6$									\\
			\bottomrule
		\end{tabular}
\end{table*}

\subsection{Gas Distribution - Pressures and Densities}
\label{subsec_results_dist}

The warm and hot gas probability distributions are shown in Figure \ref{fig:prob}. As described in \S \ref{subsec_model_cool}, the hot gas probability distribution (red curve) has a cutoff at low temperatures. This limit, $T_{\rm min}$, is set by the cooling to dynamical time ratio, one of our model parameters. In our fiducial model, hot gas cells for which $\tcool / \tdyn < 8$ form the warm component (blue curve). This is consistent with the results in \citet{Sharma12}, who find that for hot plasma in a spherical  potential, gas with $\tcool/\tdyn \lesssim 10$ can cool due to thermal instabilities and condense. $T_{\rm min}$ is calculated locally, varying as a function of radius and increasing from $0.85 \times 10^6$~K at $50$~kpc, to $1.3 \times 10^6$~K at $\rvir$.

 \begin{figure*}
\begin{tabular}{c c}
 \includegraphics[width=0.45\textwidth]{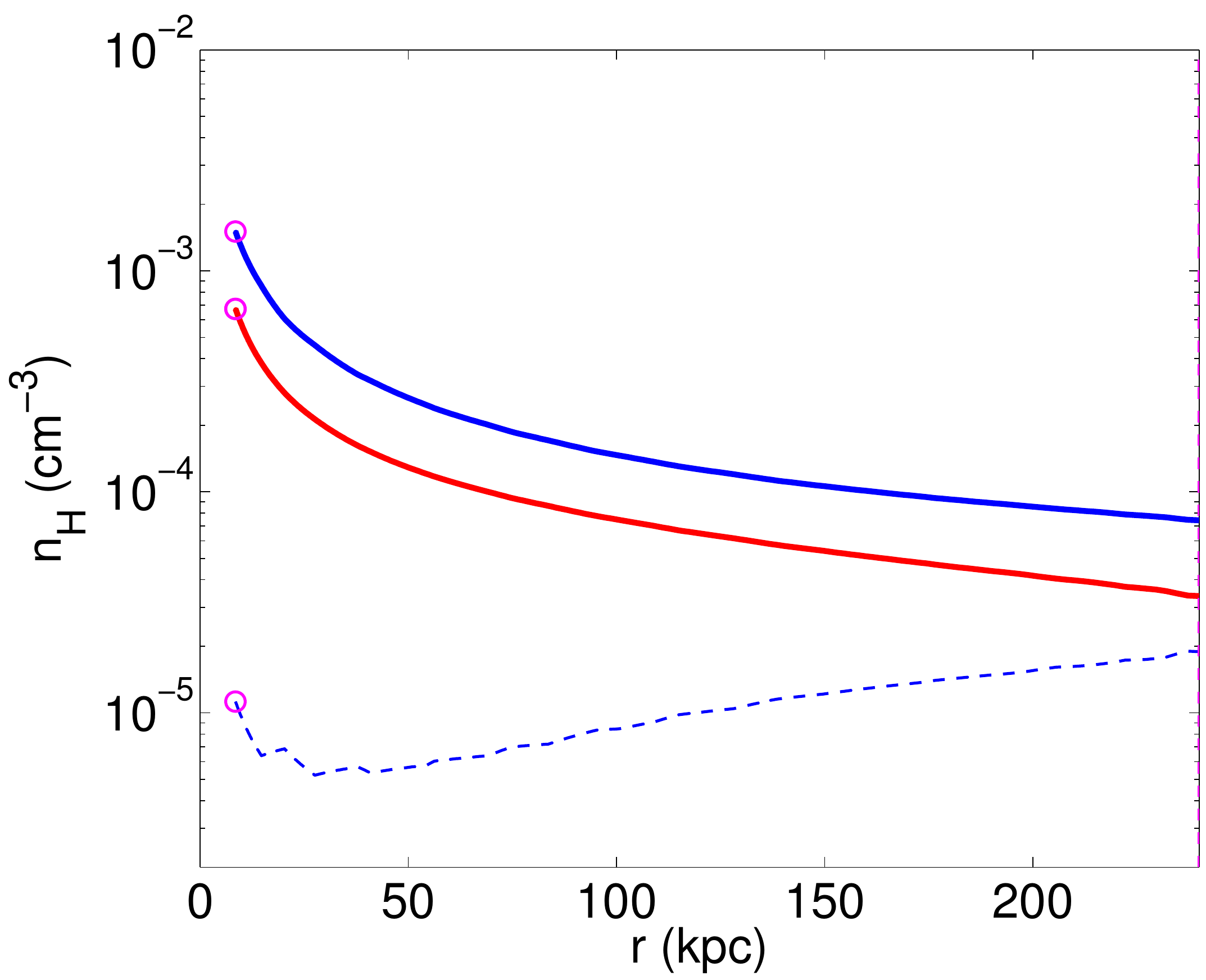} &
 \includegraphics[width=0.45\textwidth]{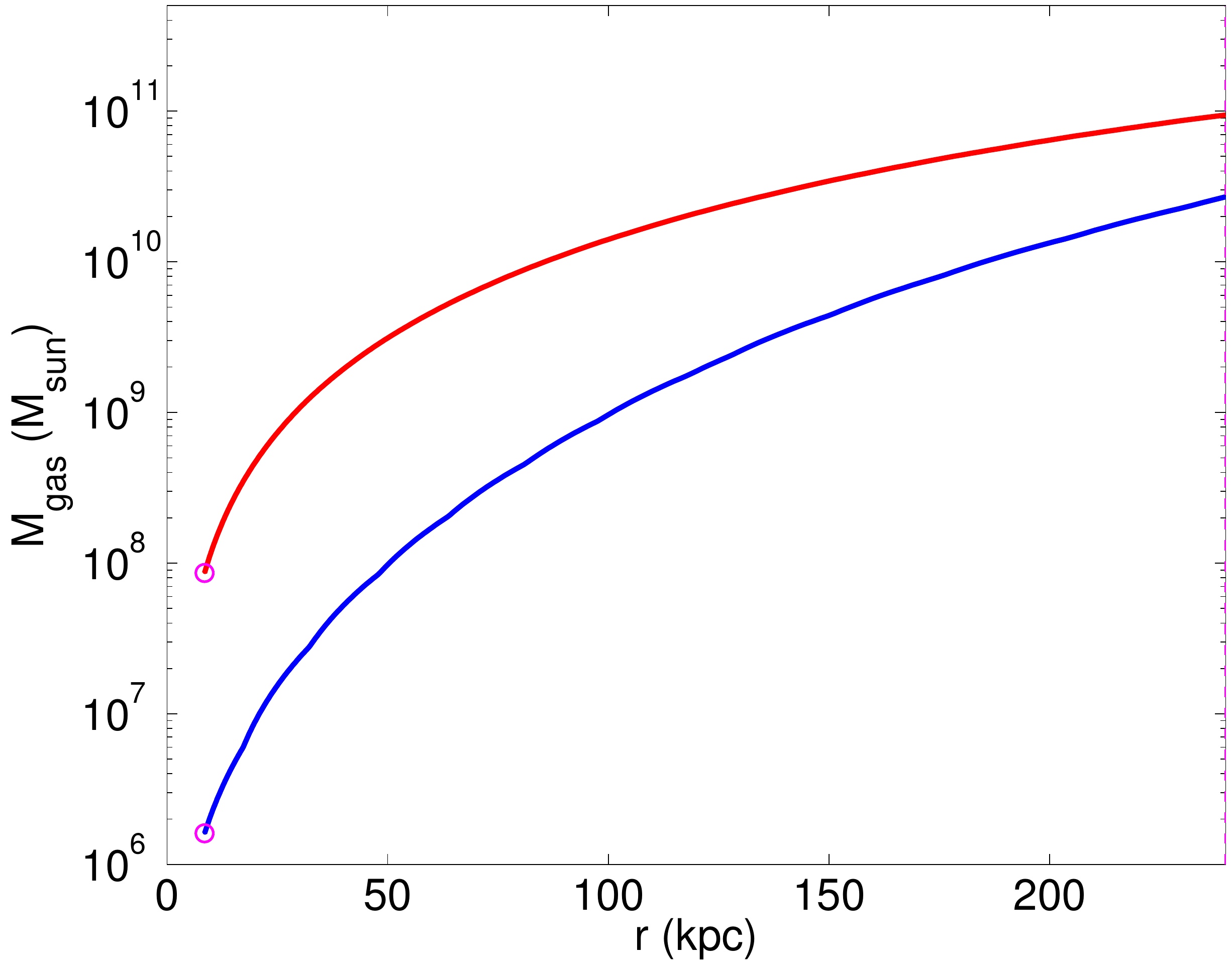} \\
\end{tabular}
\caption{Corona gas density and mass profiles for the hot (red) and warm (blue) components. 
		      {\bf Left}: The mean local density profiles (solid lines) of the hot and warm gas are extended. The hot gas is supported by the thermal, turbulent, cosmic ray and magnetic pressure. The warm gas is mixed with the hot component and follows a similar density profile. The warm gas filling factor is set by the fraction of hot gas with $\tcool / \tdyn \lesssim 8$ and inreases from $0.02$ at small radii ($\leq 50$~kpc) to $0.25$ at \rvir. The averaged density profile of warm gas (dashed blue curve) is calculated as the mean local density multiplied by the volume filling factor.  The dynamical time increases faster than the cooling time, leading to a more constant averaged density profile for the warm gas, compared to the hot and warm gas local densities. See \S\ref{subsec_model_cool} and \S\ref{sec_results} for details. The gas density profiles are attached as a data file to the online version of the paper.
			  {\bf Right}: The cumulative mass profile of the gas. The warm gas is $\sim 5\%-36\%$ of the gas mass at each radius, and the total hot to warm gas mass ratio in the corona is $\sim 3.5$. The total gas mass inside the virial radius is $1.2 \times 10^{11}$~\msun, dominating the baryonic budget of the Galaxy.}
  \label{fig:profiles}
\end{figure*}

The gas pressure profile as a function of radius is presented in Figure \ref{fig:pressure}, with the solar radius indicated by the magenta circle. The hot component dominates the gas pressure in the corona. It is supported by thermal pressure, turbulent velocity ($\sigma_{\rm turb}=60$~\kms) and cosmic ray and magnetic pressure ($\alpha - 1=0.9$, see Eq.~\ref{eq:hse}), leading to an extended gas profile. Near the virial radius, the pressure is $P/\kb \sim 230$~\cmv~K. The total ratio of non-thermal (including turbulence) to thermal support in our fiducial model is $2.1$. This is lower than the value found by \cite{OML10} in the disk, $\alpha_{\rm OML} \approx 5$, suggesting that the ratio of non-thermal to thermal pressures in the corona is diluted relative to the disk.

Extending the pressure profile to larger radii is useful for obtaining an upper limit on the pressure in the IGM. \cite{FSM13} model the HI distribution in Leo~T, a gas-rich dwarf galaxy $420$~kpc from the MW, to estimate $P/\kb \lesssim 150$~\cmv~K~at this distance. \GFS17 use a similar method and derive $P/\kb \sim 100-200$~\cmv~K~around Leo~T. Extrapolation of the pressure profile of our fiducial model gives a pressure of $\sim 100$~\cmv~K~at $r=420$~kpc. Since we expect a decrease in the gas density and pressure as we cross the virial shock (moving outwards), the pressure profile beyond the virial radius provides an upper limit on the actual IGM pressure. We can then conclude that the extrapolation to $d=420$~kpc gives an upper limit consistent with the results from \cite{FSM13} and \GFS17.

The gas density as a function of radius is shown in Figure~\ref{fig:profiles} (left panel). The mean hydrogen density, dominated by large radii, is $n_{\rm H} \sim 4.6 \times 10^{-5}$~\cmv. At smaller radii of 50-100 kpc, the density is in the range of $0.83-1.3 \times 10^{-4}$~\cmv. These densities agree with constraints derived from ram pressure stripping of MW satellites \citep{Blitz00,Grcevich09,Gatto13,Salem15} and modeling of the Magellanic Stream (see \citealp{Tepper15}). 

The density of the warm component, formed by the rapidly cooling gas, is higher than the density of the hot gas by a factor of $\sim 3$ (solid blue curve in the left panel of Figure \ref{fig:profiles}). However, the warm gas only occupies a small fraction of the total number of gas cells, with the volume filling factor given by Eq.~\ref{eq:fill_fac}. This can be seen as effective clumpiness, and the mean volume filling fraction is $ \left< \fvw \right> \sim 15\%$. The local density of the warm gas behaves similarly to the hot gas density, slowly decreasing with radius. Furthermore, the dynamical time grows with radius faster than the cooling time for a given gas cell. Thus, $\tcool/\tdyn$ decreases and more hot gas cells cool into the warm phase. As a result, the warm gas filling factor increases with radius, from $\sim 2\%$ in the central part of the corona ($\leq 40$~kpc) to $25\%$ at \rvir. This also results in an increase in the local mass fraction of the warm gas with radius, from $4\%$ to $36\%$. The warm gas averaged density profile, calculated as the product of the density and the filling factor at a given radius, is the dashed line in the left panel of Figure \ref{fig:profiles}\footnote{The hot and wam gas density profiles are attached as a data file to the online version of the paper.}.

\subsection{Oxygen columns, X-ray emission and DM}
\label{subsec_results_obs}

Our model does not include a galactic disk component. The observational data sets we use to constrain our model are described in detail in \S \ref{sec_obs}. The OVI data set we consider probes the CGM at large radii, beyond the extent of the disk. In emission, the X-ray 0.4-2.0 keV band flux and line intensity measurements we consider are corrected for local foreground and extragalactic background. The \lia~and \lib~emissions are also corrected for foreground absorption. For the dispersion measure, we use the value inferred by \cite{Anderson10}, who subtract the contribution of the Galactic disk. For the MW OVII and OVIII absorption, the observations are not corrected for any disk contributions and we use the observed values as upper limits for our model.

Figure~\ref{fig:ovi} shows the OVI column densities, comparing the model (black curve) to the observed original (blue markers) and binned data (magenta). The warm gas density profile shape and high densities result in a flat OVI column density profile, reproducing well the binned observed data. 
The mean area-weighted column density in the model is $4.6 \times 10^{14}$~\cmc. The mean of the measured column densities is $3.6 \times 10^{14}$~\cmc, with an error range of $3.3-4.0 \times 10^{14}$~\cmc. This result is affected by the low column density at $r \sim 0.4-0.5$. Excluding this radial bin brings the observed mean to $4.0~(3.6-4.3) \times 10^{14}$~\cmc, improving the agreement with the model.

The typical OVII and OVIII column densities in the model are $1.6 \times 10^{16}$ and $3.8 \times 10^{15}$~\cmc, respectively. We compare these to the typical values estimated from X-ray observations in \S \ref{subsec_obs_ovii} and presented in Table~\ref{tab:xray_obs}. For OVII we adopt the value from the F15 data set (including the non-detections), $N_{\rm OVII}=1.4~(1.0-2.0)\times 10^{16}$~\cmc. For OVIII we consider the G12 data set, with $N_{\rm OVIII}=0.36~(0.22-0.57)\times 10^{16}$~\cmc. The model column densities are consistent with the observations to within $1~\sigma$.

The total oxygen line width in our model is $b=102$~\kms. This result is consistent with the value we infer in \S \ref{subsec_obs_ovii}, of $b=98~(79-117)$~\kms. However, as discussed earlier, the absorption line width is not well determined in X-ray spectra of individual sight lines due to instrumental limitations. Thus, we can use the line width from our fiducial model to translate the individual EWs in F15 to column densities through a curve of growth calculation. The resulting observed median column density is $N_{\rm OVII} = 1.6~(1.1-2.3)$~\cmc, giving even a better agreement with our model.

\begin{figure}
 \includegraphics[width=0.45\textwidth]{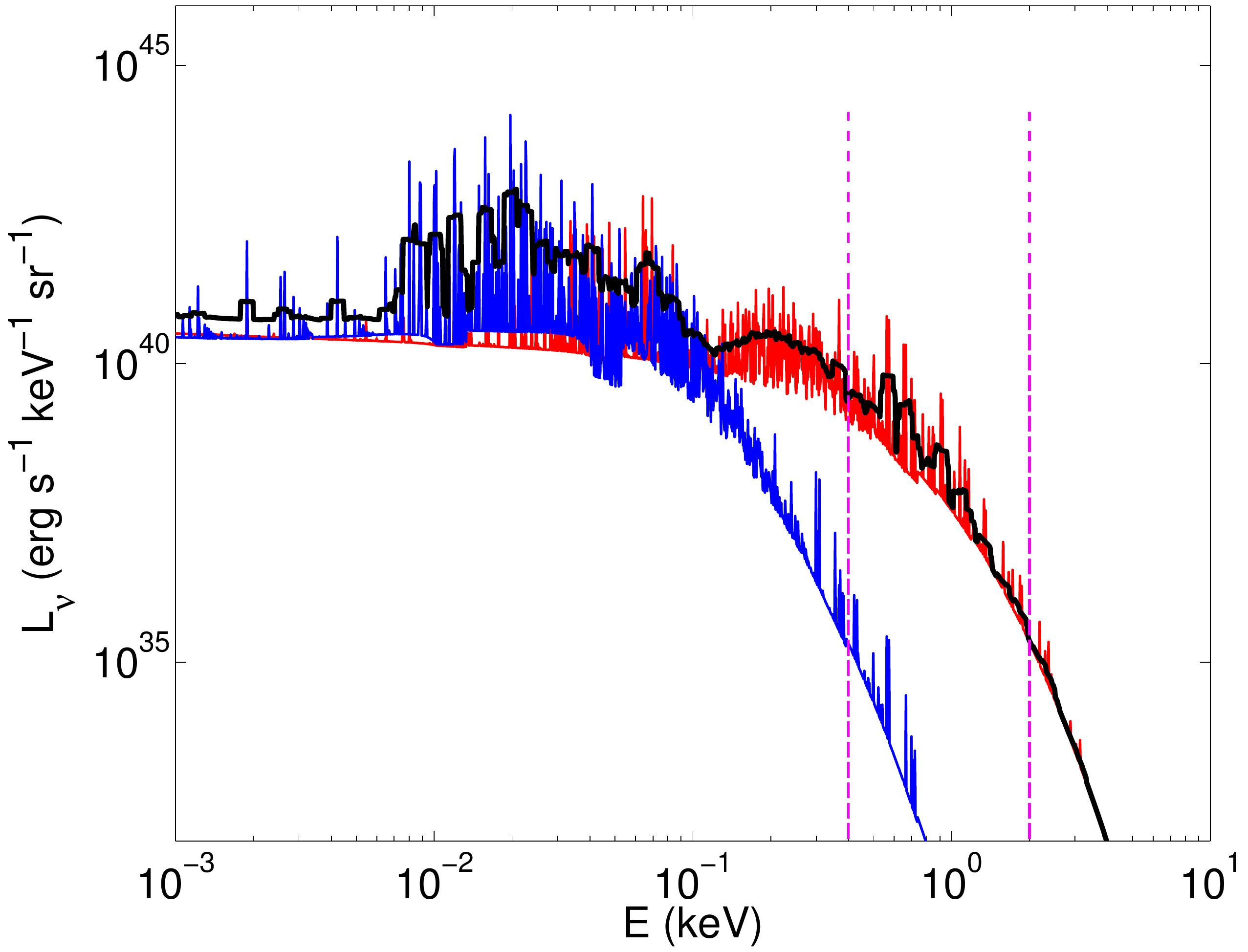}
\caption{Our predicted corona emission spectrum for the hot (red) and warm (blue) components. The luminosity in the 0.3-2 keV band (between the magenta dashed lines) is $1.5 \times 10^{39}$~\ergs and is  completely dominated by the hot component (see \S\ref{subsec_model_spectrum} and \S \ref{subsec_results_mass}).}
   \label{fig:spectrum}
 \end{figure}
 
The outer radius of the corona in our model is $\rcgm = \rvir$. This parameter affects the outer part of the OVI column density profile - for $\rcgm/\rvir<1$, $N_{\rm OVI}(h) $ decreases sharply at $h/\rvir \sim 0.8$, the outer limit of the T11 detections.  Thus, the value of \rcgm~in our model is a lower limit on the size of the corona. The OVII and OVIII total column densities, with a half column point at $\sim 50$~kpc, are not very sensitive to \rcgm.

The typical 0.4-2.0 keV flux of the corona in our fiducial model along a sightline from the Sun's location to the outer radius of the corona is $0.82$~\fluxun. This is lower than the observed value, of $2.1 (1.9-2.4)$~\fluxun. The line intensities in the \lia~and \lib~features are $1.2$ and $0.33$~\liun, respectively. These are a factor of $\sim 2$ lower than the observed values, $2.8~(2.3-3.4)$ and $0.69~(0.58-0.83)$~\liun. One possible explanation for this discrepancy is that the emission measure, dominated by the denser regions along a given line of sight, may have some contribution from hot gas in the disk. We discuss this in more detail in \S \ref{subsec_disc_emission}.

We can examine the contribution of the different transitions to the total emission in these features for our fiducial model, with a volume median temperature of $1.5 \times 10^6$~K. For the \lia~feature, $22.10~\AA$ and $21.60~\AA$ are the main transitions, contributing $\sim 45\%$ and $\sim 40\%$ of the total feature intensity, respectively. The $21.80~\AA$ transition makes up for the remainder, with $\sim 15\%$ of the intensity. For the \lib~feature, the $18.63~\AA$ OVII and the $18.97~\AA$ OVIII transitions have similar contributions.

The dispersion measure to the LMC in our fiducial model is $17.4$~\cmv~pc, consistent with the upper limit derived by \citet{Anderson10}, $DM \leq 23$~\cmv~pc. 

To compare our full model to a uniform density corona, we calculate the values of the factors \ca~and \cb~inroduced in \S \ref{subsec_mot_extent} (Eq.~\ref{eq:em} and \ref{eq:N}). In a simple constant density profile both factors have values of 1. In our fiducial model the hydrogen column density is $N_{\rm H}=8.6 \times 10^{19}$~\cmc, the emission measure is $EM=6.4 \times 10^{-3}~{\rm cm^{-6}~pc}$, and we get $\ca = 10.6$ and $\cb = 2.3$.

\subsection{Gas Masses and Radiative Losses}
\label{subsec_results_mass}

The extended and (nearly) flat gas density profile results in a large baryonic mass. The cumulative mass distribution of the gas as a function of radius is shown in Figure~\ref{fig:profiles} (right panel). The masses of the hot and warm components are  $9.4 \times 10^{10}~\msun$ and $2.7 \times 10^{10}~\msun$, respectively. The total gas mass within the virial radius is $1.2 \times 10^{11}$~\msun. Adding $M_{\rm b,obs}=6 \times 10^{10}$~\msun~(quoted in \citealt{Draine11})~results in a total Galactic baryon mass of $1.8 \times 10^{11}$~\msun. For our adopted mass profile (K02), with a virial mass of $10^{12}$~\msun, the resulting baryon fraction is $0.181$, slightly higher than the cosmological value of $0.157$. However, as discussed in \S \ref{subsec_mot_fraction}, the mass of the MW halo is uncertain to within $\sim 0.15$~dex. For a total halo mass in the range of $1-2 \times 10^{12}$~\msun, the baryon fraction in our fiducial model is in the range $0.09-0.18$. Taking the geometric mean of the mass limits, $\mvir = 1.4 \times 10^{12}$~\msun, results in a baryon fraction  of $0.128$, close to the cosmological value. Here we assume and verify that the gas mass in the model has a weak dependence on the virial mass of the halo. We explore this in more detail in paper II.

The gas loses energy through radiative cooling. The total luminosities of the hot and warm components are $5.6 \times 10^{41}$ and $7.8 \times 10^{41}$~\ergs, respectively. The emission from the two components is comparable even though the warm component mass is a factor of $\sim 3.5$ lower than the hot gas mass. This occurs because the warm gas, at temperatures around $3 \times 10^5$~K,  is near the peak of the cooling efficiency curve.  As discussed in \S \ref{subsec_model_cool}, the luminosity of the warm component is based on the assumption that the gas density remains constant in the transition from the hot to the warm component. If the transition is isobaric, the final mean gas density of the warm component will be higher by a factor $\lesssim 2.8$ (varying with radius). This will increase the cooling rate by a factor of $\sim 8$, to $\sim 6 \times 10^{42}$~\ergs, an upper limit to the radiative luminosity of the warm gas.

The emission spectrum of the corona, calculated as described in \S \ref{subsec_model_spectrum}, is presented in Figure~\ref{fig:spectrum}. The integrated luminosity of the corona in the $0.3-2.0$~keV band and inside 40 kpc is $1.5 \times 10^{39}$~\ergs, completely dominated by the hot gas. This is comparable to the typical observed value of $5 \times 10^{39}$~\ergs~(\citealt{Rasmussen09}, see \S \ref{subsec_obs_other} here).

For our fiducial model, the gas cooling times are $6.8 \times 10^{9}$ and $2.2 \times 10^8$~years for the hot and warm components, respectively. This shows that while the hot component is stable over Hubble time, the warm component will vanish on a much shorter timescale, unless an input of gas or energy is provided. As we show in \S \ref{subsec_disc_mass}, under the assumption of CIE, the short cooling time is an intrinsic property of the observed OVI, rather than a problem of our model. This result suggests that OVI coronae around galaxies are short lived and require an energy input to be stable over longer periods of time, as hinted by the non-detection of OVI around passive galaxies by T11.

%
%

\section{Discussion}
\label{sec_discussion}

\subsection{Gas Lifetime}
\label{subsec_disc_mass}

Most of the gas mass in our model ($\sim 80\%$) is in the hot component, and is a long-lived structure with a cooling time comparable to the Hubble time. This suggests that not only can hot coronae fully account for the galactic missing baryons (as suggested by \citealt{MB04, Fukugita06, Sommer06, Kaufmann06}), but they can also serve as galactic gas reservoirs,  slowly cooling, turning into warm gas, some of which can accrete onto the galaxy (see also \S \ref{subsec_disc_other} for comparison with other works). For example, at the current rate of star formation in the MW, ${\rm SFR} \sim 1.9$~\msuny~\citep{Chomiuk11}, the gas in the hot corona will be sufficient to form new stars for $\sim 50$~Gyr, suggesting that the Galaxy is in an approximate steady state.

However, the warm component, at $T \approx 3 \times 10^5$~K, is at the peak of the cooling efficiency curve and cools rapidly, with $\tcool \sim  2.2 \times 10^8$~years for isochoric cooling. We stress that the rapid cooling is not a unique result of our model, but rather a property of the OVI-bearing gas as traced by observations, under the assumption that the OVI collisionaly ionized. We can see this by following T11 and examining a simplified case of a uniform density corona with a radius of 150~kpc and a metallicity of $Z'=0.5$. At $T=3 \times 10^5$~K (where the OVI fraction is maximal, with $f_{\rm OVI}=0.2$) the cooling efficiency is $\Lambda = \mathcal{L}/n_{\rm e} n_{\rm H} = 2 \times 10^{-22}~{\rm erg~cm^3~s^{-1}}$.  The resulting cooling time, given by $\tcool \sim E_{\rm th}/L $, is $\sim 8 \times 10^8$~years, much shorter than the Hubble time. Since in CIE the OVI fraction peaks in a narrow range of temperatures and falls rapidly outside this range, a realistic value for $f_{\rm OVI}$ is significantly lower than $0.2$ and a higher total oxygen mass is needed to reproduce the observed OVI column density. For an assumed metallicity and a given volume (or corona scale length) this implies a higher gas density, $n$, increasing the cooling rate and lowering the cooling time, which scales as $n^{-1}$ for a given temperature.

Possible energy sources to balance the cooling rate of the warm component are stellar feedback processes (supernovae and stellar winds), AGN feedback, heating by satellite and IGM accretion and cosmic rays. T11 report lack of OVI absorption around galaxies with low star formation rates, suggesting that the OVI might be relatively short lived and that its energy source may be linked to stellar processes.

Given its high cooling rate, the warm gas is expected to cool to even lower temperatures. This will result in an increased density, and may cause it to sink and eventually accrete onto the galaxy, fueling star formation in the disk. However, \cite{Joung12a} perform numerical simulation of warm gas clouds in a hot ambient medium and find that these tend to be destroyed before traveling a significant distance in the corona. If the warm clouds are mixed back into the ambient medium, they may inject energy and momentum into the hot gas, heating and driving turbulence.

Another possible mechanism for creating OVI is photionization by the metagalactic radiation. As we discuss in \S \ref{subsec_disc_photo}, in our model, photionization is not significant. However, we do note that in any picture in which the OVI is photoionized (e.g. \citealp{Stern16}), the cooling problem is mitigated.


\subsection{Gas Metallicity}
\label{subsec_disc_metal}

The gas metallicity in our model, $Z'=0.5$, may be percieved as relatively high for an extended CGM. However, \cite{Suresh15} find that in the {\it Illustris} simulation galaxies with $M_\star=5 \times 10^{10}~\msun$, the CGM metallicity is $Z' \sim 0.5$ (see also \citealp{Miller16}). We emphasize that $Z' \sim 0.5$ is required in our model. Reproducing the observed high oxygen column densities with lower metallicity gas implies a higher gas mass, bringing the total galactic baryonic mass above the cosmological fraction for the MW dark matter halo. We can use this argument to derive a lower limit on the CGM metallicity. Assuming that the total corona mass that reproduces the observed oxygen column densities scales with metallicity as $Z^{-1}$, we can write $M_{\rm b,obs} +\mcor \left(0.5/Z' \right) \leq \fb \mvir$, where $\mcor$ is the corona mass in our fiducial model. This gives
\begin{equation} \label{eq:met}
Z' \geq \frac{0.5\mcor}{\fb\mvir-M_{\rm b,obs}} ~~~ ,
\end{equation}
resulting in a range of $Z' \geq 0.24-0.62$ for $\mvir = 1-2 \times 10^{12}$~\msun, $\mcor=1.2 \times 10^{11}$~\msun~and $M_{\rm b,obs}=6 \times 10^{10}$~\msun. Including the uncertainty in the observed baryonic mass of the galaxy~(${M_{\rm b,obs}} \sim 4-8 \times 10^{10}$~\msun, see \S\ref{subsec_mot_fraction}) changes this to $Z' \geq 0.22-0.79$. Lower metallicities also result in higher gas densities and thus higher cooling rates and luminosities, possibly contradicting existing constraints from X-ray emission observations.

In theory, the lower metallicity of the gas can be compensated for by extending the corona significantly beyond the virial radius, with the warm/hot gas filling a considerable fraction of the Local Group (LG) volume. While the current X-ray instrumentation cannot rule out this picture directly due to low spectral resolution, analysis by \cite{Fang06} shows that non-detections of oxygen in X-ray absorption in other galaxy groups contradict the existence of very large scale ($\sim 1$~Mpc) coronae. In addition, \cite{Bregman07} examine the correlation between the absorption strength for different sightlines and their angular offsets from the Galactic center or the LG center and conclude that the absorbing hot gas is associated with the MW.


\subsection{Emission intensities}
\label{subsec_disc_emission}

The X-ray line and band emission intensities in our fiducial model are a factor of $\sim 2$ lower than the values reported by H10 and HS10.

As described in these works (and briefly in \S \ref{subsec_obs_emission}) emission measurements are challenging, with contributions from solar-wind charge-exchange, Local Bubble hot gas and the extra-galactic X-ray background. H10 and HS10 model the measured X-ray spectra to separate the different components, and we adopt the final values they attribute to the emission from the `halo' component, corrected for local absorption. In this context, however, the emitting medium can be either disk gas or the more extended CGM which we address in our model.

For the disk contribution, \citet{Slavin00} estimate the soft X-ray emission from hot gas in supernova remnants (SNRs) in the disk, as a source of radiation photoionizing the warm ionized medium (WIM). They conclude that SNRs can account for half of the emission measure along a typical line of sight through the Galactic disk. We note that their calculation focuses on the emission around 250 eV. Extending their calculation to higher energies and comparing them to recent X-ray observations can be interesting.

We note that the emission in our model, dominated by the denser gas in the inner region of the corona, is more sensitive to the density profile shape and normalization than the column densities observed in absorption. For example, using a higher value of the thermal pressure at the solar radius, of $P_{0,{\rm th}} \sim 4000$~\cmv~K, allows us to reproduce the measured line intensities. However, since this is considerably higher than the pressure estimated from observations, even within the Galactic disk, we conclude that in the case of the MW, about half of the observed emission probably originates in the Galactic disk, as suggested by \citet{Slavin00}.

Even if half of the emission intensity we are fitting for is associated with the disk, the absorption column densities are still dominated by our extended corona. For a uniform density disk with half thickness $H$, it follows from Equations (\ref{eq:em}) and (\ref{eq:N}) that 
\begin{equation} \label{eq:disk_cont}
\frac{N_d^2}{N_c^2} = \frac{EM_d}{EM'_c} \frac{1}{Z'_c} \frac{\ca}{\cb^2} \frac{H}{L sin(b)}~~~ .
\end{equation}
In this expression the subscripts $d$ and $c$ refer to the disk and corona, $L$ is the radius of the corona and $b$ is the Galactic latitude of a given sightline. We assume that the gas disk has a solar metallicity ($Z'=1$) and oxygen abundance ($\ao = 4.9 \times 10^{-4}$) and that the disk hot gas temperature (thus, the ion fraction of a given ion) is the same as in the corona. For $H=1$~kpc and for $Z'_c=0.5$, $L=\rcgm = 250$~kpc, $\ca=10.6$ and $\cb=2.3$, as in our fiducial model, it follows that $N_d/N_c$ ranges from $0.18$ to $0.13$ for $b$ from $30^\circ$ to $90^\circ$. We conclude that the Galactic disk can produce a significant fraction of the total observed oxygen line intensity, while making only a small contribution ($\leq 15\%$) to the oxygen column density observed in absorption.


\subsection{Warm Gas Clump Size}
\label{subsec_disc_fil}

As described in \S \ref{sec_model} and \ref{sec_results}, the warm gas is mixed with the hot component but occupies only a fraction of the total volume of the corona. In an analogy to the multiphase ISM (\citealp{MO77}), we can imagine the warm gas is clumpy, with the clumps embedded in the volume filling hot gas. Assuming spherical clumps, we can estimate the typical clump size. For a clump of radius $a$, the ratio of the area fraction on the sky covered by the warm clumps, $C$, to the volume filling factor, $\fvw$ is given by $\fvw/C = a/\rcgm$, where $\rcgm$ is the radius of the corona. We take $\fvw$ to be the the mean volume filling factor, $\left<\fvw\right> \sim 0.15$ (see \S \ref{subsec_results_dist}). For $\rcgm = rvir \sim 250$~kpc and $C=1$ (see T11), we get $a \sim 36$~kpc. For an isobaric transition from hot to warm gas, the density of the warm gas is higher, the volume filling factor is decreased (for a given warm gas mass) and the clumps are smaller. Thus our estimate can be taken as an upper limit on the warm gas clump size. In addition, since the volume filling factor is a function of radius, the clump size can also vary with radius (decreasing at smaller radii).


\subsection{Photoionization?}
\label{subsec_disc_photo}

We now show that for the high gas densities of the warm component in our model, $\left< n \right> > 10^{-4}$~\cmv~(see \S \ref{subsec_results_dist} and Figure \ref{fig:profiles}), production of OVI by photoionization is not significant. \cite{GS04} calculate the metal ion densities and fractions in gas photoionized by the present-day UV/X-ray metagalactic radiation field. For a hydrogen density of $10^{-4}$~\cmv~(with $Z'=0.1)$, the photoionized OVI density is $\sim 3 \times 10^{-11}$~\cmv~(see their Model C). Correcting for the higher metallicity in our model and the updated oxygen abundance, we get $\tilde{n}_{\rm OVI} = 1.4 \times 10^{-10} $~\cmv. The gas density in the central region of our model is higher than $10^{-4}$~\cmv, and the photoionized OVI density is substantially reduced (see Figure 5 in \citealp{GS04}). Thus, we adopt $\tilde{n}_{\rm OVI} = 1.4 \times 10^{-10}$~\cmv~as the maximal attainable value via photoionization.
We can then estimate an upper limit for the photoionized column as $N_{\rm OVI} < N_{\rm OVI,max} = 2\rcgm \tilde{f}_{V,w} \tilde{n}_{\rm OVI}$, for a sightline through the center of the corona with the mean filling factor in our model, $\tilde{f}_{V,w}=0.15$. This gives a maximal column of $3.2 \times 10^{13}$~\cmc, a factor of $\sim 10$ lower than the typical column density observed by T11 and in our model (see Figure \ref{fig:ovi} and Table~\ref{tab:mod_res}). Our analysis agrees with \cite{Werk14}, who use Cloudy to show that low gas densities, of $\sim 10^{-5}$~\cmv, are required for the photoionized OVI columns to be consistent with observations (see their Figure 16, and see also \citealp{Stern16}). 

\cite{Werk13} report the detection of lower metal ions in the CGM of the COS-Halos galaxies, including MgII, SII, SiII to SiIV. In our model, the OVI is evidence for a warm component undergoing cooling from the hot gas (see also \citealp{Heckman02,Bordoloi16}). Furthermore, the radiative cooling efficiency of this gas, at $T \sim 3 \times 10^5$~K, is high and without external energy sources it will cool within $\lesssim 10^8$~years. Even in the presence of external energy sources, some of the warm gas is expected to continue cooling to lower temperatures, where photoionization could play a significant role in forming the low (MgII, CII, SiII) to intermediate (SiIV, NV) ions. In the \cite{Werk14} interpretation (see also \citealp{Stocke13}), the low ions are produced in photoionized $T \sim 10^4$~K gas with densities $\sim 10^{-3}$~\cmv. These correspond to thermal pressures of $P/k_B \sim 10$~\cmv~K, much lower than in our warm/hot corona. In our picture the cool gas giving rise to the low ions could be part of the high density tail of the OVI-bearing component that has cooled isochorically, thereby mitigating the absence of pressure equilibrium with the collisionally ionized warm/hot gas. However, this cooler photoionized component cannot contribute significantly to the OVI column density.


\subsection{NeVIII from Meiring et al. 2013}
\label{subsec_disc_neon}

\citet[hereafter M13]{Meiring13} present the discovery of 3 absorption systems at redshifts of $\sim 0.68-0.72$ in the spectrum of PG1148+549. They report the detection of ${\rm NeVIII}$ ($\lambda\lambda  ~ 770,780 ~ \AA$), ${\rm OVI}$ and several lower metal ions (CIII, OIII and OIV). M13 measure column densities of $N_{\rm NeVIII}=6-9 \times 10^{13}$~\cmc~and $N_{\rm OVI} = 0.4-2.5 \times 10^{14}$~\cmc. Furthermore, they detect a galaxy at the redshift of one of the absorbers, at an impact parameter of $217$~kpc from the QSO sightline. This detection suggests that the M13 absorbers probe the extended CGM of galaxies, similarly to the COS-Halos ${\rm OVI}$ absorbers. Motivated by their observations, we calculate the ${\rm NeVIII}$ column density profile in our model, similar to the calculation of ${\rm OVI}$, described in \S \ref{subsec_model_ocol}. At radii of $100-200$~kpc, the ${\rm NeVIII}$ column densities are $2.2-6.3 \times 10^{13}$~\cmc, comparable to the M13 measurements.
The NeVIII column density profile in our model is steeper than the OVI, and the distance of the absorber sightline from the galaxy can be a factor when comparing the model to observations. Future detections of additional absorption systems, providing a more complete sample of objects and systems will be very interesting.

M13 assume that both ions originate in the same gas phase and that the gas is isothermal. The ${\rm NeVIII/OVI}$ column density ratio can then be used to constrain the gas temperature. At CIE, the ${\rm NeVIII}$ fraction peaks at a temperature of $6.5 \times 10^5$~K, and M13 estimate $T \sim 5 \times 10^5$~K. However, in our model, most of the ${\rm NeVIII}$ column forms in the hot component. This then explains why the NeVIII density profile is steeper than the OVI (red curve versus dashed blue curve in the left panel of Fig.~\ref{fig:profiles}). This also shows that in the more realistic scenario of a non-isothermal multhi-phased corona, more ions are needed to constrain the gas properties, including temperature. In our model, the NeVIII and OVI columns depend on several parameters, among them the ratio $\tcool/\tdyn$.


\subsection{Comparison to previous works}
\label{subsec_disc_other}

Our multi-phased corona resembles the model suggested by \citet[hereafter MB04]{MB04}. In their model, the hot corona is formed by gas accretion onto the dark matter halo and shock heated to the virial temperature. Some fraction of the hot gas cools, fragments and forms the warm component, which is supported at large radii by the remaining hot gas. Some warm gas clouds can then lose angular momentum and accrete onto the galaxy, fueling star formation in the disk.

While the basic ideas of the two models are similar, there are also some significant differences. First, for the gas spatial distribution, MB04 assume that the gas follows the dark matter NFW profile; while this is not uncommon (see also \citealp{Anderson10} and \citealp{Miller13}), the physical basis for this is unclear. Second, MB04 set the gas fraction in the warm component using a threshold density and radius. Last, the metallicity in our fiducial model ($Z=0.5$) is higher than the metallicity adopted by MB04 for the coronal gas, $Z=0.1$. These differences complicate simple comparison of the two works, while detailed comparisons of corona properties for different dark matter halo masses are beyond the scope of this work.

We can compare two interesting results for the properties of the coronal gas. The first is the ratio of the cooling to the virial radius. MB04 define the cooling radius as the distance from the galactic center at which the hot gas density falls below some critical density. Thus, gas outside the cooling radius has a long cooling time and remains hot.  MB04 show that for galaxies with $V_{\rm max}\approx 120-400$~\kms, the cooling radius is smaller than the virial radius (see their \S 3.3). This is in contrast to our model, where warm gas forms at all radii of the corona, out to \rcgm. As we discuss in \S \ref{subsec_results_obs}, a cutoff in the warm gas density profile at $R<\rvir$ would be inconsistent with the observations by T11, finding significant OVI column densities at $r/\rvir \sim 1$.

The second is the warm gas fraction in the corona. In their Figure 2, MB04 show that for a halo with $V_{\rm max} \sim 200~\kms$, similar to the MW, approximately $40\%$ of the hot coronal gas mass will cool into the warm component. For halos with $V_{\rm max}$ in the range $150-250~\kms$ (bracketing the estimated value for the MW), the warm gas mass fraction is in the range of $60-25\%$.
In our fiducial model, the warm gas mass fraction is lower, around $20\%$, and it is constrained by the OVI absorption. Thus, it may serve as a lower limit on the total warm gas mass present in the corona, with additional gas at temperatures higher or lower than $3 \times 10^5$~K. However, a higher warm gas fraction poses two challenges. First, the baryon mass fraction in our model is close to the cosmological value (see \S \ref{subsec_results_mass}). Thus, a higher warm gas fraction may have to come at the expense of hot gas, with the latter constrained by the OVII and OVIII column densities. If the hot gas in the galactic disk contributes to these significantly, less hot coronal gas is required and the warm gas fraction in the corona may be higher than in our fiducial model. However, this leads to the second problem. As we have shown in \S\ref{subsec_disc_mass}, at CIE, gas at $T \sim 3 \times 10^5$~K is at the peak of the radiative coooling efficiency. A larger warm gas mass will require even more energy to balance the cooling or may lead to high accretion rates onto the Galactic disk. To summarize, while the actual warm gas fraction in the corona may be higher than the value we find for our fiducial model, $\sim 20\%$, our analysis suggests that fractions of $40-60\%$ seem less likely.

The picture in which a fraction of the warm component accretes onto the galaxy is consistent with the results by \cite{Joung12b}, who perform a hydrodynamical simulation of the CGM in a MW-sized galaxy out to the virial radius. Interestingly, they find that warm/hot gas, with temperatures in the range of $10^5-10^6$~K, dominates the accretion rates for most radii ($\sim 60-240$~kpc, see their Figure 3). They also find that the CGM mean temperature increases inwards, as the accreting gas is heated by contraction. In our analytic model, the warm component fraction decreases at smaller radii, so that the overall temperature of the gas increases. \cite{Joung12b} state that the stellar mass of their simulated galaxy, $2 \times 10^{11}$~\msun, is high compared to the MW and centrally concentrated. They do not report the total gas mass in the corona and do not present comparison to absorption or emission observations.

%
%

\section{Summary}
\label{sec_summary}

In this paper we have presented a phenomenological model for gaseous galactic corona, as traced by observations of OVI, OVII and OVIII absorption lines in the UV and X-ray, by X-ray emission observations in the Millky Way and by detections of diffuse X-ray emission around galaxies.

Previous works have usually focused just on the OVII and OVIII X-ray absorption lines. \citet{Gupta12}, for example, presented a simplified model of an extended, isothermal and uniform density corona, constrained by OVIII and reproducing the OVII column densities. While isothermal coronae at $T \sim 1-3 \times 10^6$~K can produce high OVII column densities, as demonstrated by the extended corona by \citet{Gupta12} and the cuspy, low-mass models by \citet{Anderson10} and \citet{Miller13}, these coronae do not explain the other oxygen ions, and specifically not the OVI absorption, detected by \cite{Tumlinson11}. Small scale cuspy coronae are also unable to reproduce the observed OVI spatial distribution and the density estimated from ram-pressure stripping of Milky Way satellites.

Our model (\S \ref{sec_model}) is based on simple physical assumptions. The gas is multi-phased, with warm and hot components each with a log-normal distribution in temperature and density. The warm and hot gas are each isothermal in the sense that the mean temperature is independent of position. They are turbulent with $\sigma_{\rm turb}$ taken from observations of OVI absorption in other galaxies. The hot gas is in hydrostatic equilibrium in a gravitational potential equal to that of the Milky Way and we allow for pressure support by magnetic fields and cosmic rays. The corona in our model is a large-scale structure, extending to the virial radius ($\rcgm = \rvir$). The warm phase, cooling out from the hot phase, remains mixed with it, providing a natural explanation for the extended OVI observed around star-forming galaxies in the low-redshift universe.

In \S \ref{sec_results} we presented the results of a single, fiducial model (see Table~\ref{tab:mod_par} for parameters and main properties) and compared it to the observed data, summarized in \S \ref{sec_obs}.  In our model, the OVI originates in the warm component, with a volume-weighted median temperature of $T_{{\rm med},{\mbox{\tiny {\it V}}}} = 3 \times 10^5$~K, that cooled from the hot component, with $T_{{\rm med},{\mbox{\tiny {\it V}}}} = 1.5 \times 10^6$~K and remains mixed with it out to large radii. Our model provides a good fit to the OVI column densities profile, constructed from the COS-Halos data, as presented in Figure \ref{fig:ovi}. It also produces high column densities of the OVII and OVIII ions, comparable to the observed values assuming turbulent broadening of the unresolved X-ray absorption lines (see Table~\ref{tab:mod_res}).  This is the first analytical model for a coronal structure reproducing both the UV and X-ray absorption observations, and consistent with the estimated upper limit on the dispersion measure to LMC pulsars. The X-ray emission intensities predicted by our fiducial model are lower than the observed values, by about a factor of 2 both for line and band emission. One possible explanation for this is contribution to the observed emission from the denser hot gas of SN remnants in the Galactic disk. We address this issue in more detail in paper II.

The total gas mass of the corona in our fiducial model is high, $1.2 \times 10^{11}$~\msun. Together with the long cooling time of the hot gas ($\sim 7 \times 10^{9}$~years) this suggests that hot coronae may serve as galactic gas reservoirs, capable of sustaining their current star formation rates (e.g.~$\sim 2~\msuny$ for the MW) for long periods of time. Furthermore, these massive coronae can account for the galactic missing baryons, including for the Milky Way.

The input parameters in our model are as follows: the medians and widths of the temperature distributions (hot and warm), the gas metallicity, the thermal pressure at the solar radius ($P_{\rm th,0}$), the turbulent velocity scale ($\sigma_{\rm turb}$) and the non-thermal pressure factor ($\alpha$), the ratio $\tcool/\tdyn$ defining the formation of warm gas from the turbulent hot component. The outer radius of the corona, \rcgm, is set to be equal to the virial radius. This leaves us with 9 parameters.
The observational constraints are the OVI data (column density and radial extent), the OVII and OVIII column densities, the gas density from ram pressure stripping, the dispersion measure, and the emission line intensities, with a total of 8 observables. Thus, we can only claim that our model is consistent with observations.

Several results presented in this work can be used as observational tests of the model. The first is the emission spectrum - we predict that a significant amount of the corona emission will be in the EUV, originating in the warm gas component. Second is the relatively high metallicity of our model. Furthermore, we predict a flat density profile out to large radii, which can be probed by ram-pressure constraints or by emission and dispersion measure. Last, the coronal gas can be traced using additional high ions of abundant metals, such as ${\rm NeVIII}$.

We expand this work to a full study of our model parameter space in paper II (Faerman~et~al. 2017), in which we examine how the corona properties depend on the model parameters, estimate the contribution of the hot gas in the disk to the observed absorption and emission and to what extent the uncertainties of the gas distribution or properties affect the best-fit model.

The purpose of this work is to better constrain the properties of the galactic corona by finding a range of parameters capable of fitting the observed data. We hope our models will be useful for future observations and simulations studying the large-scale galactic structure and the cycle of baryons in galaxies.

%
%

\begin{acknowledgements}

We thank Orly Gnat for assistance with our Cloudy computations of the X-ray emissivities, and for helpful discussions about this work. 
We thank Robert Braun, Aviad Cohen, Noemie Globus, David Henley, Andrey Kravtsov, Smita Mathur, Michael McCourt, Hagai Netzer, Eran Ofek and Jonathan Stern for helpful discussions about and during the course of this work. We thank the referee for the detailed comments that helped improve this manuscript.
This research was supported by a PBC Israel Science Foundation I-CORE Program 1829/12. C.F.M. is supported in part by NSF grant AST-1211729 and HST grant, HST-GO-12614.004-A. 

\end{acknowledgements}

%
%


\renewcommand{\theequation}{A-\arabic{equation}}  
\setcounter{equation}{0}  
\section*{APPENDIX}  

In our model, the coronal gas is in hydrostatic equilibrium (HSE) on large scales. On smaller scales, we adopt the results from hydrodynamics simulations, showing that local fluctuations in gas temperature and density may arise. For a given radius, we assume that these fluctuations are isobaric. On the global scale of the corona, the pressure follows the hydrostatic profile, set by the gravitational potential and the different sources of support (thermal, turbulent, cosmic rays and magnetic fields). In this Appendix we develop the relations between the properties of the local temperature distribution and the large scale spatial gas distribution, i.e. the HSE equation.

We assume a log-normal volume-weighted probability distribution for the gas temperature
\begin{equation}
g_{\mbox{\tiny {\it V}}}(x) = \frac{1}{s \sqrt{2\pi}} e^{-x^2/2s^2} ~~~,
\end{equation}
where $x \equiv \ln{(T/ T_{{\rm med},{\mbox{\tiny {\it V}}}})}$, where $T_{{\rm med},{\mbox{\tiny {\it V}}}}$ is the median temperature and
\begin{equation}
dV = V g_{{\mbox{\tiny {\it V}}}}(x) dx ~~~,
\end{equation}
for any volume element $V$.

For isobaric temperature fluctuations we now show that the
{\it mass-weighted} temperature distribution is also a log-normal.
By definition,
\begin{equation}
dM = M g_{{\mbox{\tiny {\it M}}}}(x) dx = \rho dV = \rho V g_{{\mbox{\tiny {\it V}}}}(x) dx ~~~,
\end{equation}
where $M$ is the mass within $V$, and $\rho$ is the gas density.
Thus,
\begin{equation}
g_{{\mbox{\tiny {\it M}}}}(x) = \frac{\rho}{\left<\rho\right>} g_{{\mbox{\tiny {\it V}}}}(x) ~~~,
\end{equation}
where the (local) mean density $\left<\rho\right>\equiv M/V$.  
For isobaric fluctuations in temperature we may set
\begin{equation}
\rho T = \left<\rho T\right>_{\mbox{\tiny {\it V}}} = \int \rho T g_{\mbox{\tiny {\it V}}}(x) dx = \left<\rho\right> \int T g_{\mbox{\tiny {\it M}}}(x) dx = \left<\rho\right>\left<T\right>_{\mbox{\tiny {\it M}}} ~~~,
\end{equation}
where $\left<T\right>_{\mbox{\tiny {\it M}}}$ is the mass-weighted mean temperature.  The mean $\left<T\right>_{\mbox{\tiny {\it M}}}$ enters
into the equation for hydrostatic equilibrium (HSE, see Eq.~\ref{eq:pprof}).  
Since $\rho T$ is a constant we can write
\begin{equation}
g_{\mbox{\tiny {\it M}}}(x) = \frac{1}{T\left< \frac{1}{T}\right>_{\mbox{\tiny {\it V}}}} g_{\mbox{\tiny {\it M}}}(x)  = \frac{1}{\left<e^{-x}\right>_{\mbox{\tiny {\it V}}}} g_{\mbox{\tiny {\it V}}}(x) e^{-x} ~~~.
\end{equation}

Completing the square shows that $\left<e^{-x}\right>_{\mbox{\tiny {\it V}}}=e^{-s^2/2}$ so that
\begin{equation}
g_{\mbox{\tiny {\it M}}}(x) = \frac{1}{s\sqrt{2\pi}} e^{-(x + s^2)^2/2s^2} ~~~.
\end{equation}
Thus the mass-weighted distribution $g_{_M}(x)$ is also log-normal in the temperature, 
with median $x_{{\rm med},{\mbox{\tiny {\it M}}}} \equiv \ln{(T_{{\rm med},{\mbox{\tiny {\it M}}}} / T_{{\rm med},{\mbox{\tiny {\it V}}}})} = -s^2$. Therefore
\begin{equation}{\label{eq:mvmed}}
T_{{\rm med},{\mbox{\tiny {\it M}}}} = T_{{\rm med},{\mbox{\tiny {\it V}}}} \times e^{-s^2} ~~~.
\end{equation}
For any log-normal distribution the mean and median are related by
\begin{equation}
\left<T\right> \equiv \int{T g(x) dx} = T_{\rm med} \times e^{s^2/2} ~~~.
\end{equation}
Thus, using (\ref{eq:mvmed}) it follows that
\begin{equation}
\left<T\right>_{\mbox{\tiny {\it M}}} = T_{{\rm med},{\mbox{\tiny {\it V}}}} \times e^{-s^2/2} = \left<T\right>_{\mbox{\tiny {\it V}}} \times e^{-s^2}
\end{equation}
and
\begin{equation}
T_{{\rm med},\mbox{\tiny {\it M}}} = \left<T\right>_{\mbox{\tiny {\it M}}} \times e^{-s^2/2} ~~~.
\end{equation}
The mass-weighted mean temperature that enters the HSE is smaller than the volume-weighted mean temperature 
by the factor $e^{-s^2}$ and is smaller than the volume-weighted median by the factor $e^{-s^2/2}$. 
The mass-weighted median temperature is smaller still by another factor of $e^{-s^2/2}$. 

%
%


\end{document}